\documentclass[aip,cha,reprint,superscriptaddress,nofootinbib]{revtex4-2}
\usepackage{amsmath,amssymb}
\usepackage{graphicx}
\graphicspath{{figures/}}
\usepackage{float}
\usepackage{tikz}
\usetikzlibrary{arrows.meta}
\usepackage{xcolor}
\definecolor{stblocked}{HTML}{E69F00}
\usepackage{xr-hyper}
\usepackage[colorlinks=true,linkcolor=blue!50!black,citecolor=blue!50!black,
            urlcolor=blue!50!black]{hyperref}
\usepackage{orcidlink}

\begin{document}

\title{A self-organizing mechanism for collective foraging:\\ recruitment cascades couple L\'evy search to a clustered maximum-entropy resource field}
\author{Galen J. Wilkerson\,\orcidlink{0000-0002-7957-5821}}
\email{gwilkerson@komplexai.io}
\affiliation{Komplexity AI LLC, La Jolla, CA 92037, USA}
\date{\today}

\begin{abstract}
Foraging recruitment is modeled as a contagion: an agent that finds a resource triggers a linear-threshold
cascade on the group's communication network. Because activation is degree-normalized, crowding shuts
the cascade off, so the swarm aggregates on a resource and then disperses to explore again. Acting on
the cascade moves agents in space, and that motion rewires the network, so recruitment and exploration
are one coupled process; removing this coupling alone abolishes the self-limitation. Recruitment spreads only within a connectivity window, bounded below by percolation onset
and above by degree-blocking. The swarm self-organizes to the near-critical edge of this
window---self-organized quasi-criticality: recruitment sits near but below the critical point
without settling onto a clean, size-invariant critical state, and the activity spectrum has a finite
correlation time (Lorentzian-like, not $1/f$), consistent with local rather than global criticality. Recruitment
raises harvest up to $2.7\times$ over solitary search on sparse, clustered resource fields and loses (down
to $0.37\times$) on abundant or uniform ones; because resources generically
aggregate, the favourable regime is the common one. The empirically motivated L\'evy search and the resource field it exploits are both least-committal: the
search a maximum-caliber walk, and the field a maximum-entropy point process whose generic, aggregation-set
sign is clustered (attractive). The search exponent that maximizes harvest tracks the field's fractal dimension (a heavier-tailed, more ballistic search for more strongly clustered fields).
These are one coupled mechanism: heavy-tailed L\'evy exploration crosses the field's
sparse voids while recruitment cascades exploit its clusters by switching agents to short, local steps toward the resource, so one maximum-entropy geometry binds
exploration and recruitment into a single self-organizing cycle.
\end{abstract}
\maketitle

\section{Introduction}
\label{sec:intro}

Many living systems solve the same problem: a group must find and gather rare resources scattered
unpredictably across space, using only local sensing and local communication, with no central controller and no
global map. Ants, bees, microbes, and foraging birds face it~\cite{camazine2001self,Couzin2005,sims2008}; so do engineered swarms searching for
disaster survivors, crop-pest outbreaks, or wildfire hotspots~\cite{brambilla2013swarm}. In every case the targets are typically
sparse and patchy, and the searchers numerous and short-sighted. What local rule lets such a swarm coordinate well without central
control? We study one that is biologically motivated and deliberately minimal: \emph{agents explore on their own, and an agent is recruited to a discovery once a threshold fraction of its neighbours are already active}. The rule bridges scales
in two coupled spaces: its exploration is a dispersive L\'evy search in physical
space (maximum entropy), and its recruitment spreads as an agglomerative cascade on the communication network (percolation).

\begin{figure*}[t]
  \centering
  \begin{minipage}[c]{0.46\linewidth}\centering
  \resizebox{\linewidth}{!}{%
  \begin{tikzpicture}[>={Stealth[length=2.6mm,width=2.2mm]},
      E/.style={circle,draw=black!55,fill=black!10,thick,minimum size=1.0cm,inner sep=0pt},
      R/.style={circle,draw=blue!60!black,fill=blue!12,thick,minimum size=1.0cm,inner sep=0pt},
      C/.style={circle,draw=green!45!black,fill=green!16,thick,minimum size=1.0cm,inner sep=0pt},
      D/.style={circle,draw=stblocked!80!black,fill=stblocked!30,thick,minimum size=1.0cm,inner sep=0pt},
      tl/.style={font=\normalsize,fill=white,inner sep=1pt}]
    \fill[blue!6] (4.6,1.3) rectangle (8.2,7.9);
    \draw[black!45,dotted] (4.6,1.3)--(4.6,7.9);
    \draw[black!45,dotted] (8.2,1.3)--(8.2,7.9);
    \node[font=\large\itshape,black!70] at (6.4,8.4) {cascade window};
    \node[font=\large,black] at (4.6,1.05) {$z_c^{\mathrm{RGG}}$};
    \node[font=\large,black] at (8.2,1.05) {$2/\phi$};
    \draw[->,black!70,thick] (2.4,1.3)--(11.0,1.3);
    \draw[->,black!70,thick] (2.4,1.3)--(2.4,8.8);
    \node[font=\large,black,rotate=90] at (1.85,4.9) {node activity};
    \node[E] (E) at (3.2,2.15) {};
    \node[R] (R) at (5.7,4.9) {};
    \node[C] (C) at (7.3,7.1) {};
    \node[D] (D) at (9.9,3.3) {};
    \node[font=\large\bfseries,black!75,anchor=west] at (3.78,1.7) {explore};
    \node[font=\large\bfseries,blue!60!black,anchor=east] at (R.west) {recruit};
    \node[font=\large\bfseries,green!45!black,anchor=south] at (C.north) {consume};
    \node[font=\large\bfseries,stblocked!80!black,anchor=west] at (D.east) {blocked};
    \draw[->,thick] (E) to[bend left=10] node[tl,sloped,above]{discovery} (R);
    \draw[->,thick] (R) to[bend left=8]  node[tl,sloped,above]{cascade} (C);
    \draw[->,thick] (C) to[bend left=12] node[tl,sloped,above]{block} (D);
    \draw[->,thick] (C) to[bend left=45] node[tl,pos=0.5,sloped,below]{disperse} (E);
    \draw[->,thick] (D) to[bend left=16] node[tl,sloped,below]{release} (E);
    \draw[->,black!70,line width=1.3pt] (2.7,0.25)--(4.35,0.25);
    \draw[->,black!70,line width=1.3pt] (10.7,0.25)--(8.45,0.25);
    \node[font=\normalsize,black!80,align=center,anchor=south] at (3.15,0.35) {drive\\(aggregation)};
    \node[font=\normalsize,black!80,align=center,anchor=south] at (10.05,0.35) {dissipation\\(block\,$+$\,consume)};
    \node[font=\large,black] at (6.7,-0.15) {network / spatial density  (sparse $\to$ near-critical $\to$ dense)};
  \end{tikzpicture}}
  \end{minipage}\hfill
  \begin{minipage}[c]{0.52\linewidth}\centering
    \includegraphics[width=\linewidth]{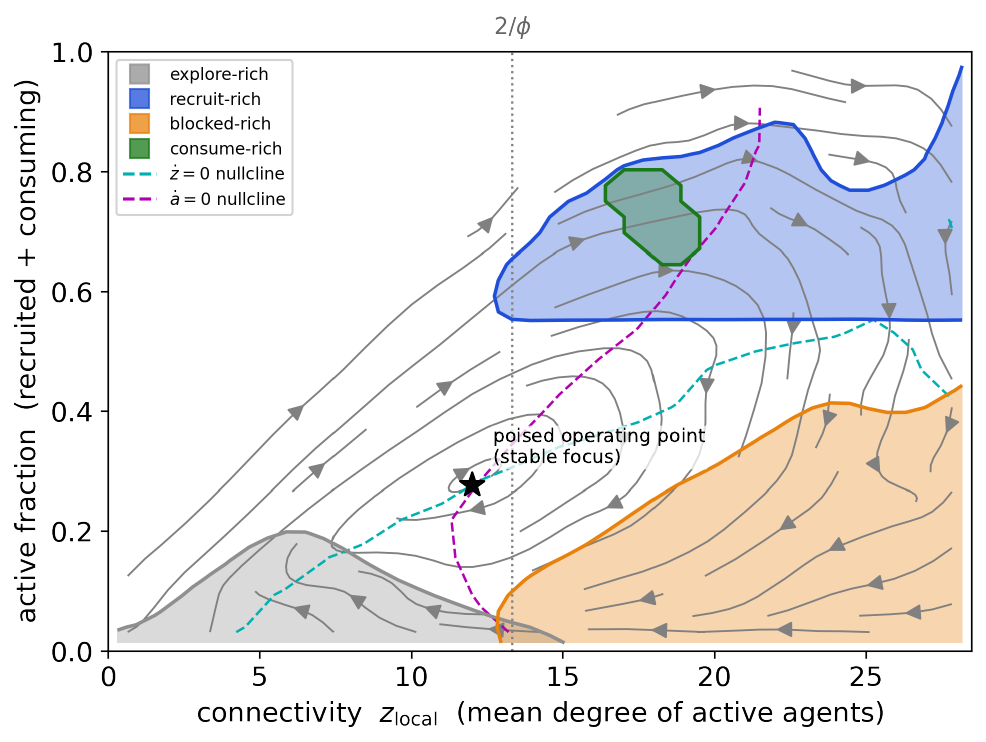}
  \end{minipage}\\[2pt]
  \makebox[0.46\linewidth]{\small\textbf{(a)} the cycle (schematic)}\hfill
  \makebox[0.52\linewidth]{\small\textbf{(b)} the cycle (measured)}
  \caption{\textbf{The swarm runs a drive--release cycle in the density$\times$activity plane:
  schematic and data.} \textbf{(a)} The cycle in the network-density$\times$activity plane. A resource
  discovery triggers a recruitment cascade (explore$\to$recruit$\to$consume); a consuming agent then
  blocks (crowded) or, once the resource is depleted, disperses to exploration, and blocked agents are
  released to explore. States: recruit (blue), consume (green), blocked (orange), explore (grey). The shaded
  band is the cascade window, bounded by the percolation onset $z_c^{\rm RGG}$ and the blocking onset
  $2/\phi$ ($\phi$ the activation threshold; \S\ref{sec:results-q1}); inward arrows show the swarm driven
  into the window from both sides. \textbf{(b)} Empirical phase portrait (128 replicas): shaded regions
  mark where each state is richest (bin-mean occupancy $\ge1.5\times$ its plane-average; Methods), with
  grey mean-flow streamlines circulating around the swirl centre (star), and the $\dot z{=}0$ (cyan) /
  $\dot A{=}0$ (magenta) drift nullclines, which cross at the drift fixed point (star, $A{\approx}0.28$: where the
mean flow vanishes, not the swarm's time-averaged active fraction; cf.\ \S\ref{sec:results-q3}). The measured circulation
  realizes the schematic cycle; the operating point is approached by an overdamped, excitable relaxation.}
  \label{fig:cycle}
\end{figure*}

Played out over time, the rule produces a foraging cycle (Fig.~\ref{fig:cycle}; in physical space,
Fig.~\ref{fig:overview}): an explorer discovering a resource triggers a recruitment cascade over its
spatial network; recruits switch to short local moves and converge on the resource; crowding then blocks
further recruitment, and agents revert to long-hop exploration.

The loop has a fixed causal order. The resource field is exogenous and, because resources generically
aggregate, sparse and clustered (\S\ref{sec:generic}); the agents do not reshape it. Its two features set
the two search modes: long L\'evy hops cross the voids, and local convergence exploits a cluster once
recruitment finds one (Fig.~\ref{fig:overview}). Movement and recruitment then close a loop
(Fig.~\ref{fig:cycle}a): a cascade pulls explorers into convergence, whose crowding reshapes the network
the next cascade travels on. With neither search nor recruitment fine-tuned, one minimal rule should coordinate
effective foraging across environments.

\begin{figure}[t]
  \centering
  \includegraphics[width=\linewidth]{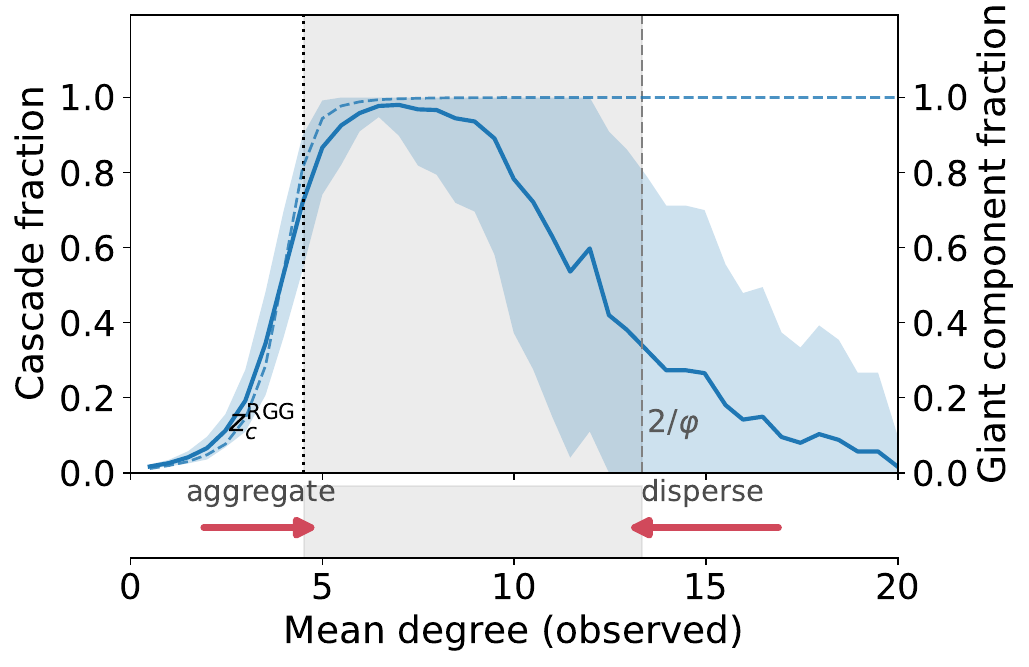}
  \caption{\textbf{Recruitment works only in a connectivity window (too sparse to spread below,
  self-blocked above), and the swarm self-tunes into it.} Mean cascade fraction ($\pm1$ s.d.) versus observed mean degree $z$ for the
  linear-threshold rule on a proximity graph. Recruitment turns on near the random-geometric-graph
  percolation threshold $z_c^{\mathrm{RGG}}\approx4.51$ (giant-component fraction, dashed, right axis)
  and is suppressed again at higher density (blocking), peaking at intermediate connectivity: two
  boundaries from the single fraction rule $a/k\ge\phi$. The giant component visibly grows
  below $z_c^{\mathrm{RGG}}$ through finite-size smearing (finite-size scaling places the
  transition at $z_c^{\mathrm{RGG}}$; Methods). The red arrows below the axis are the restoring
  connectivity flow, the sign of $\dot z$ from the cycle (Fig.~\ref{fig:cycle}): too sparse, cascades
  aggregate agents and raise $z$; too crowded, blocking disperses agents and lowers $z$; so the population self-tunes
  into this cascade window. Quantified in \S\ref{sec:results-q1}.}
  \label{fig:invertedU}
\end{figure}

We begin from a property of the \emph{linear threshold model}
(LTM)---a simple network cascade model of social contagion---not from a task. In the LTM each node stays inactive until the fraction of its neighbours
that are active reaches its threshold~$\phi$, whereupon it activates and can trigger
others~\cite{Granovetter1978,Watts2002}. Because activation is on the fraction $a/k\ge\phi$ ($a$ active neighbours out of degree $k$), cascade
size is non-monotonic in connectivity (Fig.~\ref{fig:invertedU}). The lower boundary is a percolation transition~\cite{Watts2002}: below it
the cluster of vulnerable nodes, those a single active neighbour can activate, does not span the
graph, so cascades stay local; once that cluster percolates, a discovery can recruit a system-spanning
fraction. But as degree grows further the same fraction rule renders high-degree nodes non-vulnerable, and
cascades are blocked. This non-monotonic response of cascade size to connectivity is a control surface
(Fig.~\ref{fig:invertedU}), whose two boundaries drive the cycle the swarm traces (Fig.~\ref{fig:cycle}).

This control surface is static---it treats connectivity as a fixed input. But the nodes are mobile agents
whose recruitment draws them together, increasing connectivity, so it becomes endogenous (\S\ref{sec:model}). We take
collective foraging, the biological archetype of this spatial task, as our worked example.

\textbf{Research questions.} Five questions test that claim---that one untuned rule makes the swarm an effective forager across the generic regime.
\textbf{Q1 (structure)} Does spatial linear-threshold recruitment exhibit a cascade window---an
inverted-U recruitment response to connectivity? \textbf{Q2 (operating point)} Where in that window does
the swarm settle under its own dynamics, and is the branching local or global? \textbf{Q3 (mechanism)}
Is the self-limitation caused by the recruitment--motion coupling? \textbf{Q4 (efficiency)} How efficient
is the coordinated regime, in harvest, movement, and communication, and what sets each? \textbf{Q5
(generality)} The mechanism pays off only on sparse, clustered fields (Q4); is that field the generic
one, and does the maximum-entropy search match it?

\section{Related work}
\label{sec:related}
\textbf{Active matter.} Like active matter, our rule produces collective order from purely local
interaction, so it connects to the flocking literature. It does so by a different route: where Vicsek-type
models order through velocity alignment~\cite{Vicsek1995,Toner1995,vicsek2012collective,bowick2022active}, order
here is recruitment-mediated (a discrete cascade, not alignment), task-driven, and
transient (agents aggregate at a resource, deplete it, disperse). We reproduce collective order
and then add a mechanism.

\textbf{Threshold cascades and percolation.} Watts'
LTM on random graphs undergoes a percolation-type onset once a giant component of vulnerable nodes
percolates~\cite{Watts2002} (continuum percolation of the underlying random geometric
graph~\cite{Penrose2003,mertens2012continuum,DallChristensen2002}); the rule was introduced for
social contagion, adoption and rumour spread~\cite{Granovetter1978,Watts2002}, a setting now studied on
time-varying graphs as well~\cite{unicomb2021cascades}, which recruitment
here spatially embeds. We depart twice: the graph is induced by agent positions, which recruitment then reshapes by
drawing those agents together (endogenous connectivity), and we use both boundaries, percolation below and
degree-blocking above (the mobile analogue of rumour stifling~\cite{daleykendall1965,makithompson1973}), as a productive window rather than
studying the onset alone.

\textbf{Spatial cascades.} Embedding an LTM on a random geometric graph rather
than a non-spatial one makes recruitment edge-efficient: a small connection radius, and so few
links per node, still spreads large, system-spanning cascades~\cite{wilkerson2021logic}; we turn that
static spatial cascade into a mobile one. The network cascade is the channel by which a local
discovery becomes a spatial rearrangement across the swarm.

\textbf{Swarm robotics / task allocation.} Response-threshold
models switch task when a stimulus crosses an individual
threshold~\cite{brambilla2013swarm,bonabeau1999swarm,gordon2010ant,gordon1996regulation,bonabeau1996quantitative,dorigo2021swarm};
ours thresholds the fraction of active neighbours, and it is this normalisation that yields
self-limitation for free, where engineered controllers add explicit inhibition. The explore--exploit balance this strikes is the province of reinforcement learning, formalised by the multi-armed bandit~\cite{robbins1952bandit,lattimore2020bandit,SuttonBarto2018}; a companion paper develops the connection~\cite{wilkerson2026learnerfree}.

\textbf{Foraging theory.}
Optimal foraging characterises the single searcher: the \emph{marginal value theorem} (leave a patch as
its intake falls to the habitat average~\cite{charnov1976optimal}) and L\'evy random
search~\cite{viswanathan2011foraging,Viswanathan1996}, whose inverse-square optimality for $d\ge2$ remains
debated~\cite{levernier2020inverse,buldyrev2021comment,levernier2021reply,carames2022levy}. Here these
single-searcher behaviours emerge together from one local rule run by a collective, the patch-leaving
a qualitative correspondence and the L\'evy optimum an empirical one (Table~\ref{tab:coherence}).

\textbf{Self-organized criticality and excitable media.} Our swarm sits in the non-conserving
self-organized-criticality family: like the Bak--Tang--Wiesenfeld
sandpile~\cite{bak1987soc,jensen1998soc,BeggsPlenz2003,obyrne2022criticality,cavagna2023swarms} it is
slowly driven and relaxes through recruitment avalanches, and like the non-conservative
Olami--Feder--Christensen earthquake~\cite{olami1992soc} and Drossel--Schwabl forest-fire~\cite{drossel1992forest}
models its blocking dissipates activity in the bulk. Such non-conserving systems self-organize not to the critical
point but to its neighbourhood---\emph{self-organized quasi-criticality}~\cite{bonachela2009soqc}, hovering
near criticality without settling onto a clean critical point---which is the regime we find
(\S\ref{sec:excitable}). The dynamics are those of a driven, dissipative excitable
medium~\cite{meron1992excitable} on a mobile substrate, where the avalanche moves the nodes themselves.
Against these prior pieces---Watts' cascade and its percolation onset~\cite{Watts2002}, 2D continuum
percolation~\cite{DallChristensen2002}, non-conservative SOC~\cite{olami1992soc,drossel1992forest},
spatially embedded adaptive networks~\cite{gross2008adaptive,berner2023adaptive}, L\'evy
search~\cite{viswanathan2011foraging}, and our own static spatial cascade~\cite{wilkerson2021logic}---what is
new is their combination on a mobile substrate: the three departures above (endogenous connectivity, both
boundaries as one window, and the resulting self-organized quasi-critical cycle) make the foraging
behaviours follow from one coupling rather than being added by hand.

\section{Model}
\label{sec:model}
We study the foraging cycle of \S\ref{sec:intro} on a 2D resource field (Fig.~\ref{fig:overview}). Its
coordination is a compact primitive, defined independent of environment and of the search mode: (i) an
interaction network topology $G_t$ induced by proximity (a \emph{random geometric graph}, RGG~\cite{Penrose2003}), (ii) a binary
fraction-threshold recruitment rule on $G_t$, which cascades when the active-neighbour fraction is high and
intrinsically self-limits (\emph{blocking}) as neighbourhoods densify, and (iii) a coupling from activation
to motion. The exploration law (L\'evy here, Brownian equally) and the resource field are layers this
primitive couples to, not part of it.

\textbf{State and dynamics.} A square torus $[0,L)^2$ (periodic boundaries, to avoid edge effects and
keep every location statistically equivalent) holds $N$ agents and $M$ localized resources
(minimum-image distances). Inactive agents explore by heavy-tailed L\'evy flights (stable index
$\alpha$, $0<\alpha<2$; step-length tail exponent $\mu=1+\alpha$); an agent overlapping a resource becomes a seed and consumes it; depleted resources respawn
elsewhere, keeping the field statistically stationary. Agents $i,j$ share an
edge when $d(\mathbf{x}_i,\mathbf{x}_j)\le r$ (a random geometric graph), rewired as agents move. Because
information propagates faster than agents, the LTM rule is iterated to its fixed point each tick: with
$a_i$ active neighbours and degree $k_i$, each agent's binary state $s_i\in\{0,1\}$ ($1{=}$active) updates asynchronously as
\begin{equation}
  s_i \;\longleftarrow\;
  \begin{cases}
    1, & k_i>0 \ \text{ and }\ a_i/k_i \ge \phi,\\[2pt]
    s_i, & \text{otherwise,}
  \end{cases}
  \label{eq:ltm}
\end{equation}
activating monotonically to the Watts global-cascade fixed point. Recruited agents take a short local step (the exploitation scale, not the exploration step length), directed
toward the centroid of their recruiters (aggregation solves the task); seeds do local area-restricted search. Convergence
raises local density, so $k_i$ grows, $a_i/k_i$ falls below $\phi$, and recruitment is
blocked, releasing agents. This topology--geometry feedback (cascades reshape geometry, which
rewires the graph, which regulates percolation and blocking) is the central mechanism. Per-tick pipeline:
move $\to$ seed $\to$ cascade to fixed point (Eq.~\ref{eq:ltm}) $\to$ converge $\to$ commit.

\section{Methods}
\label{sec:methods}
\textbf{Notation.} Controls: $\phi$ activation threshold (rule: active-neighbour fraction
$a_i/k_i\ge\phi$), $\rho_r=N(r/L)^2$ connectivity density, $\rho_s=M(\ell/L)^2$ resource coverage
($\ell$ patch size; abundance), $\alpha$ L\'evy stable index (step-length tail exponent $\mu{=}1{+}\alpha$), $\delta\phi$ threshold heterogeneity. Per agent: $k_i$ degree,
$a_i$ active-neighbour count, $s_i$ state; $A$ = population active fraction. Connectivity: $z=\pi
p\rho_r$ mean degree ($p$ link probability), $z_{\rm local}$ over active agents, $z_c^{\rm RGG}$ percolation onset. Criticality:
$\sigma$ branching ratio, $\chi$ susceptibility, $S$ avalanche size, $\Psi$ largest-cluster fraction.
Capitals denote population aggregates; lowercase-with-subscript, per agent.
Dispersion is reported as s.d./SEM.

\textbf{Dimensionless controls.} Measured against $L$ and one tick, the model's raw parameters
($N, r, M, \ell, \phi$) reduce to three genuine dimensionless controls: $\rho_r$
(propagation/percolation), $\rho_s$ (seeding), and the cascade threshold $\phi$ (recruitment vs.\
blocking), a data collapse in which distinct raw settings then behave identically. Threshold
heterogeneity $\delta\phi$ is a further control, held at zero in the baseline and characterised in
\S\ref{sec:results-cost}. Phase diagrams in $(\rho_r,\phi)$ at fixed $\rho_s$ are therefore statements
about a class of systems, independent of box size or agent count.

\textbf{Simulation.} We simulate the model of \S\ref{sec:model} at concrete scale: $N\sim10^2$ agents,
horizons of order $10^3$ ticks, batches of a few tens of independent replicas (each a fresh random realisation, with new agent positions,
graph, and seed nodes). Every observable is a
function of the dimensionless controls (above) and is itself dimensionless or intensive
(a fraction, ratio, or per-agent rate), so results describe the class of systems, not the
particular $N$; consistently, the cascade window's critical point is $N$-independent across
$N\in\{50,\dots,3200\}$ (the SI). Recruiting and
no-recruitment conditions (the latter with $\phi$ raised so no cascade spreads) are compared
paired within the same randomised context; each reported quantity is an ensemble average over a
batch of independent replicas (batch size per figure).

\textbf{Uncertainty.} Where shown, uncertainty is over these
replicas: shaded bands are $\pm1$ s.d.\ or $\pm1$ SEM (as labelled) for curves, error bars likewise for
points, $95\%$ bootstrap confidence intervals for scalar estimates (correlations, fitted exponents), and
heat-map cells are batch means.

\textbf{Cost ledgers.} Each run tracks four
dimensionless per-agent-tick cost ledgers, namely movement, presence-beacon pings (the once-per-agent-tick
local broadcast by which agents within range $r$ discover one another and form the graph $G_t$),
cascade-activation pings (one message per newly recruited agent as the signal of Eq.~\ref{eq:ltm}
propagates), and harvest, kept separate and never summed, so no efficiency metric is baked into the dynamics; cost
is otherwise reported as work to complete a fixed job (consume a set resource fraction), in
steps/target and bits/target.

\textbf{Exploration exponent.} Two L\'evy settings appear and should not be conflated: harvest and
efficiency results (Q4; Fig.~\ref{fig:phasediagram}) use $\alpha{=}1$, the
empirically observed forager optimum (\S\ref{sec:results-cost}), whereas the criticality diagnostics
(Q2; $\sigma_{\rm local}$ and the susceptibility, Figs.~\ref{fig:recruitcluster},~\ref{fig:susceptibility})
use $\alpha{=}1.5$. Because the near-criticality is a property of the cascade dynamics, not of how agents move (\S\ref{sec:excitable}), the
particular exploration law does not affect the Q2 conclusions.

\textbf{Connectivity caveat.} Mean degree $z$ is a convenient control, but two things stop it from
fixing behaviour by itself. First, $z=\pi p\rho_r$ holds exactly only for uniform positions: the swarm
aggregates, so realised degree exceeds the formula, and the same $z$ reached with a different link
probability $p$ gives a different realised degree ($p{=}1$ vs.\ $0.25$ differ by ${\approx}2\times$). We
therefore hold $p{=}1$ and vary connectivity through the density $\rho_r$. Second, $z$ is topological and
ignores the physical travel scale set by $r$: topological statistics collapse onto $z$, but harvest and
efficiency vary substantially with $r$, so $z$ alone does not fix efficiency.

\textbf{High-dimensional efficiency map.} The full control space (14 dimensionless parameters) is sampled
with a 4096-point Sobol low-discrepancy design~\cite{sobol1967} (a space-filling quasi-random sequence that covers the
space far more evenly than random draws) rather than a grid ($4^{14}$ cells is infeasible);
nuisance parameters are randomised rather than frozen, recovering each marginal at linear cost.

\textbf{Estimators.} Each Results quantity is measured as follows. Local branching
$\sigma_{\rm local}$ is the branching-process (Galton--Watson) offspring-to-parent ratio within a single
cascade ($\sigma_{\rm local}{=}1$ marks criticality~\cite{BeggsPlenz2003}). This conditioned estimator is biased
upward near criticality, however: conditioning on cascades that actually spread over-counts offspring. Applied to synthetic Galton--Watson cascades under the same
observed-cascade conditioning it returns ${\approx}1.0$ across true $\sigma\in[0.7,1.0]$, so
$\sigma_{\rm local}{\approx}0.9$ is an upper-leaning estimate rather than a sharp point value, and the
near-critical reading rests also on the system-spanning cascades and the spectral signature
(\S\ref{sec:excitable}). The cluster-size susceptibility is
$\chi=\langle S^2\rangle/\langle S\rangle$ over recruited-cluster sizes $S$ (excluding the largest cluster,
as usual in percolation); for finite-size scaling we instead use the order-parameter susceptibility
$\chi=N\,\mathrm{Var}(\Psi)$, with $\Psi$ the largest-active-cluster fraction, swept over
$N\in\{50,\dots,3200\}$. Giving-up density~\cite{brown1988patch} (the marginal-value signature~\cite{charnov1976optimal}) is the resource remaining
when a patch's occupancy collapses from its peak.

\textbf{Cascade-size finite-size scaling (Fig.~\ref{fig:fsscollapse}).} We record every cascade's
connected-component size and sweep the swarm size $N\in\{50,\dots,800\}$ with all dimensionless controls
held fixed ($r{=}L\sqrt{\rho_r/N}$ auto-rescales; source count $\propto N$). To compare fields at
criticality we calibrate the activation threshold per $(N,\text{field})$ so $\sigma_{\rm local}{\approx}1$
(bisection), then test the collapse $s^{\tau-1}P(S{\ge}s)$ vs $s/N$. On the near-CSR field (sparse, close to complete spatial randomness) the tuned cascades form a near-critical band (median $\tau{\approx}1.56$; no clean size collapse); strongly clustered fields give a lower, geometry-set
exponent, and degenerate point-clumping cannot be brought to $\sigma_{\rm local}{=}1$ at all. Confidence
intervals are bootstrapped over the independent replicas.

\textbf{Recruitment reach (Fig.~\ref{fig:frontspeed}).} From an instrumented run we log,
each tick, agent positions and activation state. The recruitment reach is the (periodic) distance
from each newly activated agent to the nearest agent active the previous tick, that is, how far coordination
propagated across space in one step; the agent displacement is the per-tick move length. We compare
the two distributions over all ticks past burn-in (the initial transient discarded).

\textbf{Phase portrait (Fig.~\ref{fig:cycle}b).} Pooling all replicas over $t>t_{\rm burn}$, every sample
is binned in the (connectivity $z_{\rm local}$, active fraction $A$) plane and the per-node state fractions
(explore\slash recruit\slash consume\slash blocked, defined in \S\ref{sec:excitable}) averaged within each bin; the binned
mean drift $(\langle\dot z\rangle,\langle\dot A\rangle)$ gives the streamlines and the $\dot z{=}0$,
$\dot A{=}0$ nullclines, whose crossing (a fixed point of the mean flow, into which the streamlines
flow) marks the stable fixed point. A state's shaded ``rich'' region is an enrichment contour of
its (smoothed) mean-fraction field: bins where that state's mean occupancy is at least $1.5\times$ its
plane-average, the same factor for all four states (prevalence-aware, so the common explore state and the
rare consume state are treated alike), a visualization of where each state concentrates, not a sharp
phase boundary. The consume state is rare, so its region is drawn as the largest contiguous enriched
component, dilated slightly for legibility, rather than the scattered cells the bare threshold yields.

\section{Results}
\label{sec:results}
The mechanism is untuned: its ingredients are least-committal rather than fitted. We take the four
mechanism questions in turn (Q1--Q4), then ask whether the regime where it pays off is generic enough for the
mechanism to recur (Q5); simulation and estimator details
are in \S\ref{sec:methods}.

\textbf{Terminology.} We reserve \emph{critical} for the cascade-window transition itself (universality
class established separately at a \emph{tuned} critical point, \S\ref{sec:results-cost}); the swarm's own
untuned operating point self-organizes not \emph{to} that transition but to its near-critical neighbourhood
(\emph{self-organized quasi-criticality}~\cite{bonachela2009soqc}, \S\ref{sec:results-q2}). ``Marginal'' we
keep for its foraging meaning alone (\S\ref{sec:related}).

\subsection{The cascade window: percolation and blocking (Q1)}
\label{sec:results-q1}
Two opposing effects act along the connectivity axis. Reach: below a connectivity threshold the interaction
graph is fragmented and cascades stay local; near the giant-component onset a discovery recruits
system-wide. For a 2D random geometric graph this occurs at a critical mean degree
$z_c^{\mathrm{RGG}}\approx4.51$~\cite{DallChristensen2002,mertens2012continuum}, a property of RGG
connectivity, independent of the task, setting the window's lower boundary. Blocking: because
activation is on the active-neighbour fraction $a/k$, rising degree lowers it below the threshold $\phi$ and suppresses propagation. Watts'
upper boundary makes high-degree nodes non-vulnerable at $k>1/\phi$ (a single active neighbour then gives fraction $1/k<\phi$)~\cite{Watts2002}. Their competition
yields the non-monotonic (inverted-U) response of Fig.~\ref{fig:invertedU}: cascade fraction rises near
$z_c^{\mathrm{RGG}}$ and peaks near $z{\approx}7.0$ (peak ${\approx}0.98$) before the blocking falloff.
What is new is that connectivity is set by the agents' own
positions, so the swarm moves within this window as it aggregates. Consistently, the mean degree
inside cascaded clusters sits at an approximate blocking boundary $2/\phi$ (e.g.\ ${\approx}10$ at
$\phi{=}0.20$); approximate because it is a Watts-vulnerability estimate: the per-node boundary is
$1/\phi$, but the mean-degree onset lands near $2/\phi$ because a node is typically recruited once
${\approx}2$ of its neighbours are active, so $a/k$ first crosses $\phi$ near $k\approx2/\phi$.

\subsection{The swarm self-organizes to quasi-criticality (Q2)}
\label{sec:results-q2}
Under the resource-search task the swarm rides a self-limiting explore/exploit cycle. The global
branching ratio (secondary activations per activation, $\sigma{=}1$ at criticality) stays deeply
subcritical (mean $\sigma\approx0.03$): the average is dominated by exploring agents, and, atypically for
a spreading process, adding connectivity drives $\sigma$ \emph{down} past the cascade-window peak, because higher degree blocks
propagation. Resolved per avalanche the branching is higher, $\sigma_{\rm local}\approx0.9$ (on the
swarm's own L\'evy exploration, $\alpha{=}1.5$), and it rises as the threshold loosens ($0.83$ at
$\phi{=}0.35$ to $0.99$ at $\phi{=}0.20$), consistent with blocking setting how far below the transition
the operating point sits. Because this estimator is biased upward (\S\ref{sec:methods}), $\sigma_{\rm local}\approx0.9$ places the operating point
near the transition; that it sits below, not at, follows from the finite-size non-collapse
and the activity spectrum (\S\ref{sec:excitable}), not the estimator.

\begin{figure*}[t]\centering
  \includegraphics[width=\linewidth]{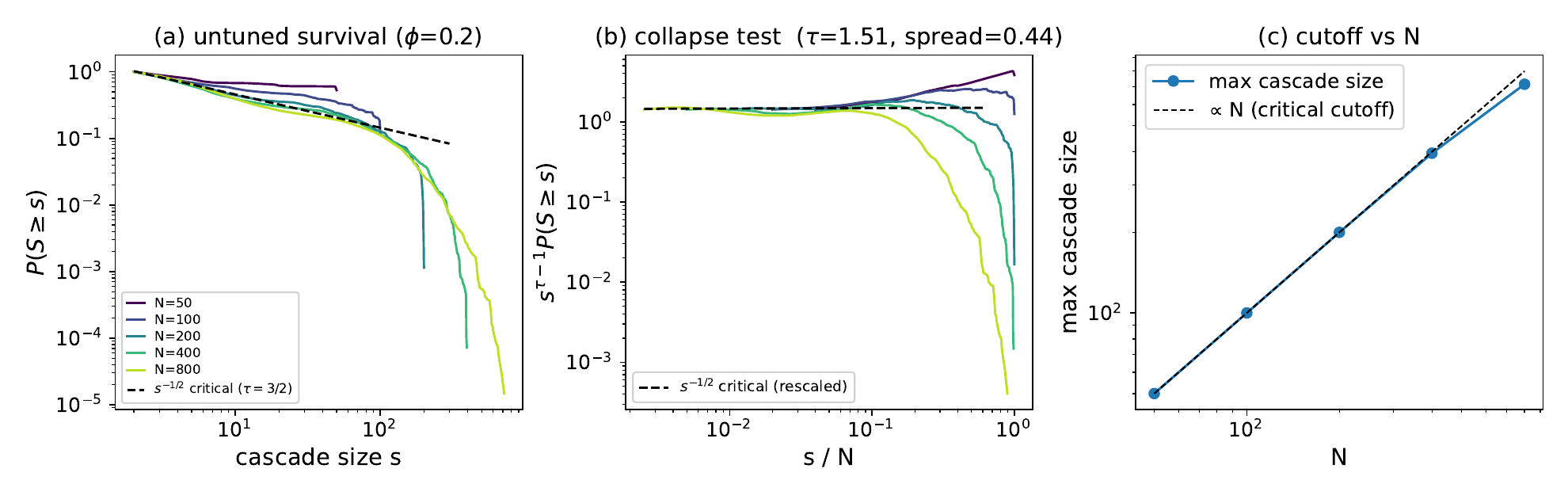}
  \caption{\textbf{Untuned finite-size test of cascade sizes: near-critical but no clean collapse.} Fixed-threshold
  ($\phi$ untuned) cascade-size distributions across $N\in\{50,\dots,800\}$. \textbf{(a)} Survival
  distributions; the dashed line is the mean-field critical slope $s^{-1/2}$ ($\tau{=}3/2$), which the
  curves track through the scaling region before rolling off. \textbf{(b)} Rescaled $s^{\tau-1}P(S{\ge}s)$ vs $s/N$ overlay in the scaling region (a power-law body
  near $\tau{=}3/2$) and separate only at the finite-size cutoff---not a clean, size-invariant collapse of
  the full distribution (log-spread $0.44$, above the ${<}0.15$ of a good collapse). \textbf{(c)} The largest cascade spans essentially the whole swarm at every size
  (maximum ${\approx}N$, tracking the $\propto N$ reference, black dashed), which rules out subcriticality.
  System-spanning cascades, with the branching estimator near but below $\sigma{=}1$ and no universal
  collapse (panel b), mark self-organized quasi-criticality---near-critical but not a size-invariant
  critical point---not true conservative SOC.}
  \label{fig:untunedcollapse}
\end{figure*}

A finite-size test locates the operating point. Holding $\phi$ fixed (untuned) and sweeping
$N\in\{50,\dots,800\}$, two things hold together (Fig.~\ref{fig:untunedcollapse}). First, the largest
cascade spans essentially the whole swarm at every size (maximum ${\approx}N$), while the mean cascade
stays small (tens of agents): the operating point is not strongly subcritical---a strongly subcritical process
would almost never span the system---so it sits near the transition, which the upward-biased per-avalanche
estimator alone could not establish. Correspondingly the branching estimator converges to
$\sigma\approx0.96$ at the largest $N$ (down from $1.03$ at $N{=}50$)---near but below $1$. Second, the
rescaled cascade-size distributions overlay in the scaling region---a power-law body near the mean-field
slope $\tau\approx3/2$---but separate at the finite-size cutoff, which pulls inward with $N$; the full
curves do not collapse onto a single master curve (log-spread $0.44$, against the ${<}0.15$ of a clean
collapse). Taken with the system-spanning
maxima and sub-unity branching ratio above, the untuned operating point is near-critical but not a
clean, size-invariant critical point---self-organized quasi-criticality~\cite{bonachela2009soqc}, placing
the swarm with the non-conserving Olami--Feder--Christensen and forest-fire models rather than at a true
(conservative) SOC fixed point. Even at the separately tuned transition
(\S\ref{sec:results-cost}) the cascades remain a near-critical band, not a clean mean-field collapse (the SI). System-spanning recruitment is
what gives the swarm its dynamic range: most finds trigger small responses, while an occasional one sweeps
a large fraction of the population.

\textbf{Even the tuned transition is a near-critical band.} At a separately \emph{tuned} transition
($\sigma_{\rm aval}{\approx}1$, matched across $N\in\{50,\dots,800\}$) on the near-CSR field, the
cascade-size distributions still do not collapse onto a single clean curve but fall in a band
(rescaled by the median $\tau{\approx}1.56$; Fig.~\ref{fig:fsscollapse}, the SI); strongly clustered fields give a lower exponent. So even deliberate tuning does not reach a clean, size-invariant critical state;
the size-independent critical point is pinned instead by the order-parameter susceptibility (the SI).

\subsection{The recruitment$\to$motion coupling causes self-limitation (Q3)}
\label{sec:results-q3}
\begin{figure*}[t]
  \centering
  \includegraphics[width=\linewidth]{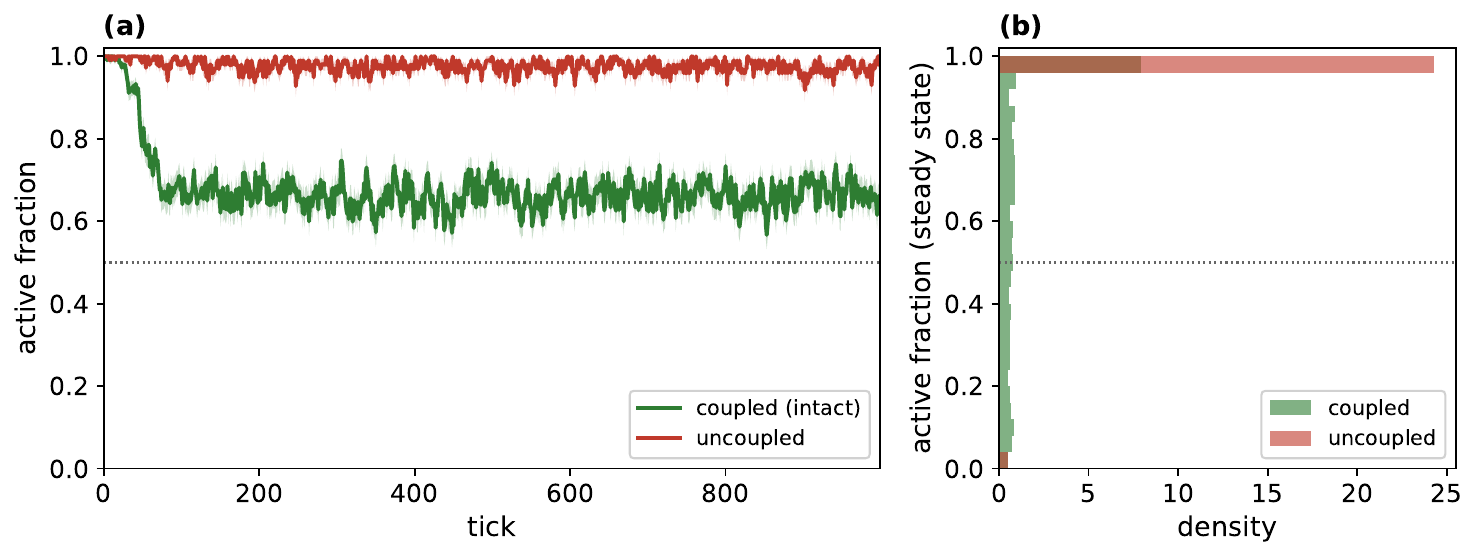}
  \caption{\textbf{The recruitment$\to$motion coupling causes the self-limitation: a controlled
  comparison.} Only the coupling is toggled. \textbf{(a)} Active fraction $A$ over time (mean $\pm$SEM): with the
  coupling (green) the swarm self-limits around half active; uncoupled (red), activation floods and
  saturates near $1$. \textbf{(b)} Steady-state distribution ($t>200$): coupled occupies a broad
  sub-saturation band, uncoupled piles against $1.0$. Removing one link, not any parameter, abolishes the
  balance.}
  \label{fig:q3ablation}
\end{figure*}

\looseness=-1 To separate cause from correlation, we run a controlled intervention: hold every parameter fixed and toggle the
recruitment$\to$motion coupling. With it intact the swarm self-limits to ${\approx}0.66$ active
(riding the cycle; this time-averaged active fraction is a different quantity from the active fraction
$A{\approx}0.28$ at the mean-flow fixed point of Fig.~\ref{fig:cycle}b---the crossing of the pooled drift
nullclines in the $z_{\rm local}$--$A$ plane, not a time average); removing the coupling, so that agents still activate (the cascade still
spreads) but no longer converge on the resource, makes
activity saturate toward ${\approx}100\%$ (Fig.~\ref{fig:q3ablation}). The mechanism is direct: without
convergence recruits never crowd a resource, so the driver for local degree $k$ to climb is removed: the
activated fraction $a/k$ never falls below the threshold $\phi$, and blocking never occurs. The same uncoupling collapses the resource--exploration correlation to ${\approx}0$ (\S\ref{sec:excitable}). Because only one link is cut while all else is untouched, this isolates cause, not association.

\textbf{Recruitment reaches across the swarm in one cascade.} A single cascade recruits agents far
beyond the step an agent takes in a tick. The median distance from a newly activated agent to the nearest
already-active one is ${\approx}193$ world units---about $126\times$ the ${\approx}1.53$ units an agent
moves per tick (Fig.~\ref{fig:frontspeed}); active clusters span most of the arena. This is a per-cascade recruitment
reach against a locomotion step, not a physical front speed: the quantity shows that coordination travels
along the percolating network, which spans distances an agent would cross only over many ticks. Reach is
carried by the network, not by locomotion.

\subsection{Efficiency: what recruitment buys, and what it costs (Q4)}
\label{sec:results-q4}\label{sec:results-cost}
\begin{figure}[t]
  \centering
  \includegraphics[width=\linewidth]{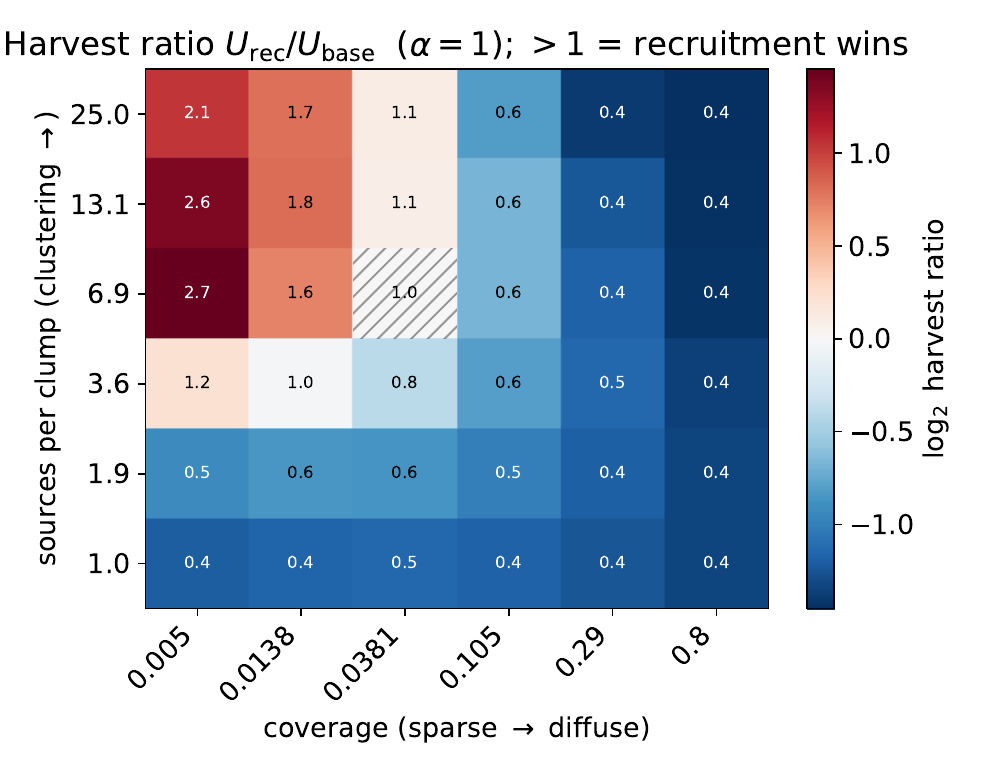}
  \caption{\textbf{Recruitment beats solitary search only for sparse, clustered targets.} Harvest
  ratio (recruitment harvest divided by the no-recruitment baseline) over patchiness (sources per clump)
  and coverage $\rho_s$: ${>}1$ (warm) means recruitment wins, ${<}1$ (cool) means it loses, with cells
  hatched where the $95\%$ CI includes $1$ (indistinguishable from baseline). Recruitment pays off for sparse,
  clustered fields and loses when targets are uniform or abundant. The gain falls with coverage and broadly (not strictly) with patchiness, from $2.7\times$ (sparse, clustered; $95\%$ within-cell replica CI $[2.71,2.77]$) to $0.37\times$ (abundant,
  diffuse), dropping below $1$ in the abundant corner. Computed at L\'evy
  stable index $\alpha{=}1$; a full sweep shows that $\alpha$, how efficient solitary search already is,
  decides whether recruitment repays its cost (\S\ref{sec:results-cost}).}
  \label{fig:phasediagram}
\end{figure}

\paragraph{Harvest: where recruitment pays.} Whether recruitment pays off depends entirely on how resources are distributed. We sweep patchiness (sources per clump) and coverage $\rho_s$ against a no-recruitment baseline, where the harvest ratio is the number of resources collected with recruitment divided by the number collected without, over the same number of time steps (Fig.~\ref{fig:phasediagram}): in abundant or uniform fields recruitment over-concentrates, with agents
piling onto found sources, cutting harvest to as low as ${\sim}0.37\times$ the no-recruitment baseline while adding movement and messages, so
it loses even before communication is priced. In sparse, clumped fields it pays off (at $\alpha{=}1$, harvest up to
$2.7\times$ the no-recruitment baseline, $95\%$ within-cell replica-bootstrap CI $[2.71,2.77]$; larger when solitary search is inefficient, \S\ref{sec:results-cost}): a discovery is informative, so recruiting sends agents to a rich neighbourhood for
cheap local exploitation, and blocking makes agents leave a patch before it is emptied (emergent
patch-leaving, \S\ref{sec:foraging}), so their effort is not wasted over-exploiting a depleting patch but
returned to search. Over a 36-cell $\alpha{=}1$ grid, recruitment wins in the sparse, clustered region ($8/18$ clumped cells vs.\
$1/18$ diffuse), losing where resources are abundant or diffuse, a clean
boundary over the resource statistics, and exactly where real foraging does not operate
(patchy resources are the default).

\paragraph{Costs: movement and communication.}
Exploration and exploitation call for different motion, and the cascade selects it: exploring agents take
long-range L\'evy steps (advantageous on sparse, clustered targets, and breaking the shared oversampling of a cohort
released from one patch), recruited agents take short local steps. The heavy tail thus does double
duty: the least-committal search matched to a scale-free field (\S\ref{sec:maxent}), and the long-range
escape that carries a released cohort off a drained patch rather than re-crowding it. Two things
distinguish this from single-walker L\'evy theory, which supplies optimality but not a
trigger~\cite{viswanathan2011foraging}: the switch is driven by a social signal (a
neighbour's discovery via the cascade), and the case for long jumps is itself social.

A full sweep over the 14-parameter space separates recruitment's two cost channels. Movement it
\emph{lowers}: against an identical non-recruiting swarm, a recruited agent reaches each target in $0.12$--$0.57\times$ the
non-recruiting steps (roughly $2$--$8\times$ fewer, the saving largest at small $\alpha$---long-range
search), because it switches to short local exploitation rather than ranging on. In
absolute terms the recruiting swarm spends only a few steps per target; the non-recruiting baseline runs
from a few to ${\sim}20$, U-shaped in $\alpha$ (cheapest near $\alpha{\approx}1$). The harvest gain itself grows as solitary search worsens---from
${\sim}2.7\times$ at $\alpha{=}1$ toward an order of magnitude by $\alpha{\to}2$ (Brownian). One caveat: the exploitation step length is itself a strong
lever on the harvest ratio---a property of the motion model, not the resource field---so we hold it fixed
for the phase diagram. The communication premium is the opposite sign: a ${\sim}5\times$ message cost on the activation channel,
governed mainly by connectivity and threshold ($z,\phi$) rather than $\alpha$ (a mild ${\sim}30\%$ drift), so cost and payoff are set
by different controls, and the message cost can be tuned through connectivity without disturbing
the regime that decides whether recruitment pays off. (Channel-by-channel numbers, and the finite-size scaling that
pins the window at a size-independent $z^\ast\approx7.8$ with peak $\propto N$, are in the SI.) Threshold heterogeneity $\delta\phi$ raises
cascade size (recruited fraction $0.50\to0.61$) but leaves efficiency and the
cascade exponent unchanged---a separate axis from the
resource field, so the homogeneous baseline is a genuine simplification, not a favourable special case.

\paragraph{Empirical grounding.} Both ingredients, the forager's heavy-tailed search and the sparse,
scale-free resource fields it exploits, are documented, and, crucially, independently. Diverse
marine predators move by long-range L\'evy search with heavy tails (step-length exponent
$\mu\approx2$ in $P(\Delta)\propto\Delta^{-\mu}$ ($\Delta$ the step length), i.e.\ stable index $\alpha=\mu-1\approx1$, the value used
here, so our baseline sits at the empirically observed value)~\cite{sims2008} (though strict power-law claims in movement data warrant care over visual fits~\cite{stumpf2012critical}); measured resource fields are sparse and scale-free (Kalahari canopy
clusters, cluster-size exponent $\beta_{\rm res}<2$~\cite{scanlon2007}, rainforest gaps~\cite{solemanrubia1995}, turbulent
fields~\cite{sreenivasan1997turbulence}). The forager exponent ($\mu\approx2$) and the field exponents
($\beta_{\rm res}<2$) were measured by different investigators studying different quantities, with no fit tying
them; that a heavy-tailed search exponent and the fields it searches are mutually compatible is a
falsifiable fact, not a definition (\S\ref{sec:maxent} argues it is no coincidence). Driving the model
with genuinely scale-free fields confirms recruitment still pays off ($2.7$--$4.1\times$ the baseline, more on
more-clustered fields), and the cascades stay near-critical there, but with a field-set exponent: on these clustered fields $\tau_{\rm active}\approx1.2$, below
the near-CSR value (median $\tau\approx1.5$, itself a near-critical band, \S\ref{sec:results-q2}), because the cascade sizes then carry
the field's patch geometry rather than the branching dynamics---the swarm self-organizes to
near-criticality on any such field, but the exponent is the field's, not a constant it imposes. Heavy-tailed search is meanwhile advantageous on
clustered targets (the SI), the regime beyond the scattered (Poisson) targets for which inverse-square L\'evy is provably not optimal in $d\ge2$~\cite{levernier2020inverse}.

\subsection{The regime where recruitment pays off is generic, and the search matches it (Q5)}
\label{sec:generic}
The mechanism pays off conditionally: on sparse, clustered fields it wins (up to $2.7\times$ at $\alpha{=}1$,
\S\ref{sec:results-q4}); on uniform or abundant ones it loses. Whether that makes it general or a
special-case trick turns on two questions: is a sparse, clustered field generic, and does the
least-committal search match it?

The first question---is a sparse, clustered field generic?---turns on the \emph{sign} of the field's
spatial correlation: aggregation or regularity. Attraction is the generic sign. The least-committal (maximum-entropy) field at a given mean clumping is a
Gibbs point process $P(\text{field})\propto e^{\beta\mathcal{C}}$~\cite{moller2004statistical,baddeley1995area}
($\mathcal{C}$ close pairs; the SI, Fig.~\ref{fig:entropy}), whose sign fixes the arrangement: regular
($\beta{<}0$), random ($\beta{=}0$), clustered ($\beta{>}0$); this field-interaction $\beta$ is distinct from the clump-size exponent $\beta_{\rm res}$. Independent processes all drive $\beta{>}0$ and
compound---offspring beside parents, autocatalytic renewal on a shared
substrate~\cite{rietkerk2004self}, and physical concentration by turbulent
transport~\cite{sreenivasan1997turbulence}. Regularity alone needs an active dispersive mechanism and exact
randomness is a measure-zero knife-edge, so non-aggregation is what takes a mechanism: the clustered field is
the generic one, and the one the mechanism needs. The argument is combinatorial and physical, so it carries
to engineered target fields too---maximum entropy fixes the family, renewal physics fixes the sign, and the
environment sets only the magnitude of $\beta$.

Sparsity sets that magnitude where it matters. The effective clustering strength
$\beta_{\rm eff}\approx\log(\langle\mathcal{C}\rangle/\langle\mathcal{C}\rangle_{\rm CSR})$ grows
logarithmically as the field thins (within-clump close pairs ${\sim}N\mu$ against a CSR denominator
${\sim}N^2$; the SI, Fig.~\ref{fig:betasparsity}: fitted slope $0.91$), so the sparsity foraging presupposes
sharpens the very structure recruitment exploits. Measured fields bear this out: forest-gap and
canopy-cluster power laws~\cite{sprott2002forest,scanlon2007} and turbulent
intermittency~\cite{sreenivasan1997turbulence} are not merely clustered but strongly, often scale-free
clustered, so the large-$\beta$ regime is what the data report.

The payoff settles the second. The gain rises with clustering, below one at the
near-CSR field and up to ${\sim}2.7\times$ at strong aggregation (Fig.~\ref{fig:phasediagram}; the harvest
ratio tracks the effective clustering $\beta_{\rm eff}$ introduced above), and drops below one
only for abundant, diffuse fields a sparse-field forager never meets. So the least-committal search matches
the field in practice---recruitment exploits exactly the clustered structure, and loses only where foragers
do not go.

\begin{figure*}[t]\centering
  \includegraphics[width=0.92\linewidth]{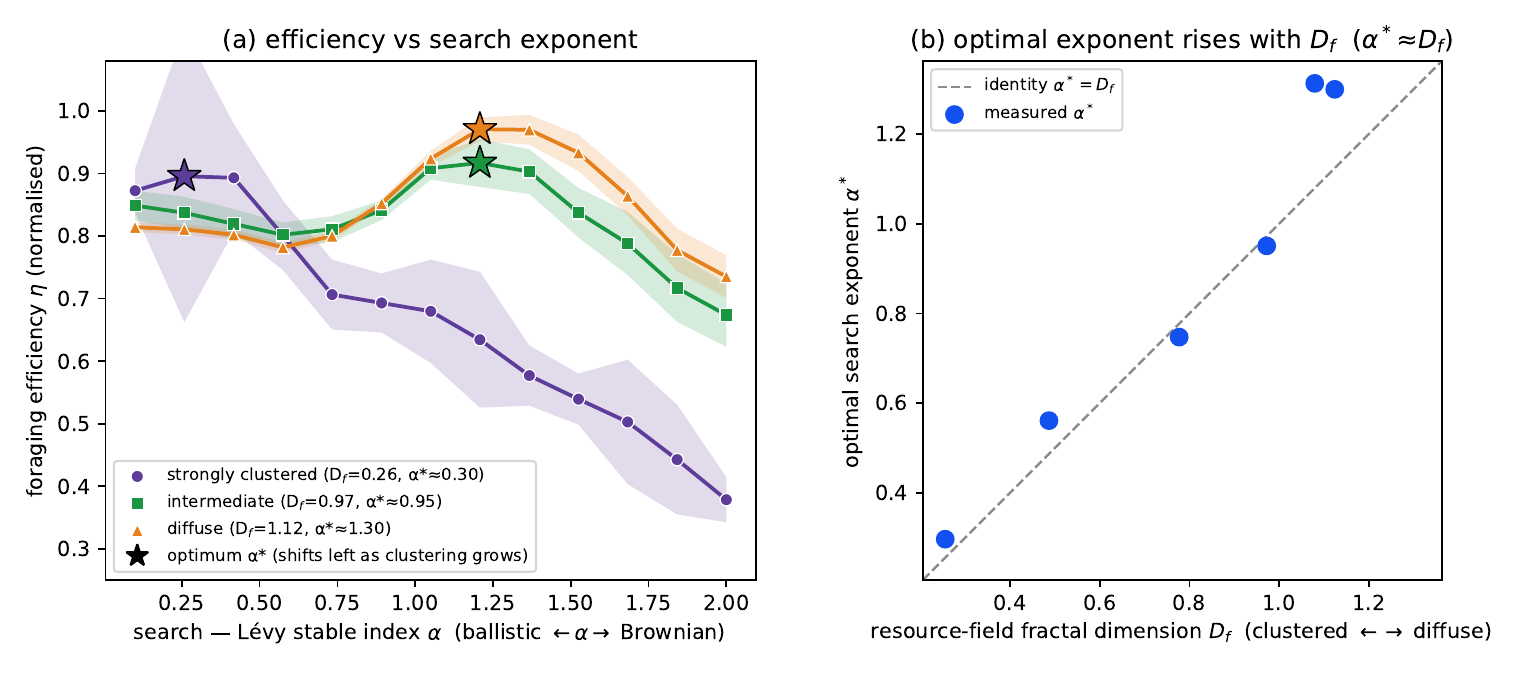}
  \caption{\textbf{The optimal search exponent tracks the field's clustering (a recruitment-free
  single-searcher sweep).} Destructive foraging (targets per unit path length) on L\'evy-dust fields
  spanning fractal dimension $D_f\!\in\![0.26,1.12]$, swept over the L\'evy stable index $\alpha$ (tail
  exponent $\mu{=}1{+}\alpha$; small $\alpha=$ heavier tail). \textbf{(a)} Efficiency $\eta(\alpha)$ for
  three representative fields ($\pm$SEM bands): strongly clustered fields peak ballistic
  ($\alpha\!\to\!0$), diffuse fields peak near inverse-square ($\alpha\!\approx\!1.2$); the Brownian end
  ($\alpha\!\to\!2$) is worst throughout. \textbf{(b)} The optimum $\alpha^\ast$ rises monotonically with
  $D_f$, $\alpha^\ast\!\approx\!D_f$ over the clustered range and saturating near inverse-square for diffuse
  fields ($r{=}0.97$ over six fields, 95\% CI $[0.87,0.99]$, $192$ samples/point; raw-argmax $r{\approx}0.86$). The ends
  recover known limits (inverse-square for diffuse targets~\cite{viswanathan2011foraging}, ballistic for
  strong clustering); the new content is the monotonic interpolation. This is a numerical tracking, not a
  proven identity, and is the quantitative form of the compatibility \S\ref{sec:maxent} argues.}
  \label{fig:caliberopt}
\end{figure*}

The search matches the field by the same principle. Refusing to privilege any step scale gives scale-free,
heavy-tailed L\'evy exploration (its maximum-caliber basis is \S\ref{sec:maxent}), matched to a field with
no scale either; the field's two features map onto a two-mode search, L\'evy crossing the sparsity and
recruitment exploiting the clustering. The match is quantitative: the optimal search exponent tracks the
field's fractal dimension, $\alpha^\ast\!\approx\!D_f$ (Fig.~\ref{fig:caliberopt}).

\section{Discussion}
\label{sec:discussion}
The Results established a self-limiting recruitment mechanism that pays off on sparse, clustered fields
(\S\ref{sec:generic}). It appears here as a driven excitable cycle and a foraging strategy, and as a
reinforcement-learning policy in a companion paper~\cite{wilkerson2026learnerfree}; across these readings the coordination lives in the
coupling of a percolating recruitment network to the physical space the agents search.

\subsection{A driven, self-limiting excitable medium}
\label{sec:excitable}
Read dynamically, the rule is a driven, self-limiting excitable cycle on a moving proximity graph
(Fig.~\ref{fig:cycle}): the slow drive of exploration and respawn is punctuated by cascade avalanches,
with blocking as the refractory step that prevents runaway~\cite{meron1992excitable}. The measured
phase portrait realizes this cycle as an overdamped relaxation to a stable fixed point
(Fig.~\ref{fig:cycle}b). Blocking produces an emergent explore/exploit balance, the exploring fraction
self-balancing rather than tuned---the coupling-caused self-limitation established in Q3. It is
transport-independent: the swarm self-limits whether agents explore by L\'evy or Brownian walks, because
it arises from recruitment and blocking, not from how agents move (L\'evy's separate contribution is
foraging efficiency on scale-free fields, \S\ref{sec:maxent}).

This self-limiting cycle is that of the non-conserving SOC models (\S\ref{sec:results-q2}): the
quasi-critical operating point is local (per cascade), while the swarm as a whole is globally subcritical
(mean $\sigma\approx0.03$, diluted by the exploring majority). We test this local-not-global reading with
the activity power spectrum.

\begin{figure}[tbp]\centering
  \includegraphics[width=\linewidth]{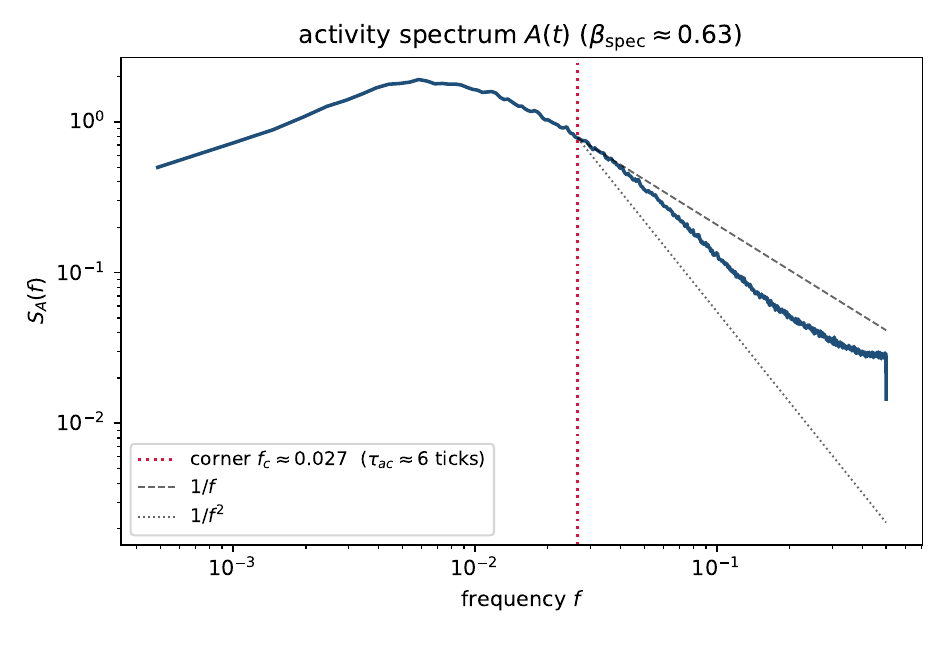}
  \caption{\textbf{The recruitment activity spectrum has a finite correlation time (Lorentzian-like, not $1/f$).}
  Power spectrum of the global recruitment activity $A(t)$ (active fraction, sampled per
  tick; instantaneous, not integrated, which would add a spurious $f^{-2}$), at the swarm's operating point
  in the stationary replenished task ($\rho_r{=}2$, so $z_{\rm local}\approx7.4$ over the $\approx16\%$
  active agents; $\phi{=}0.20$, $\alpha{=}1.5$; $192$ replicas, $16{,}384$ ticks after a $2000$-tick
  transient), computed on the canonical cascade engine, which replenishes sources so that $A(t)$ is
  stationary. $S_A(f)$ (replica mean) is a flat plateau below a corner $f_c\approx0.03$ (crimson, set by the activity
  autocorrelation time $\tau_{\rm ac}\approx6$ ticks, marked), then a roll-off between the $1/f$ and $1/f^2$
  guides---not the scale-free $1/f$ of a long-memory, globally critical process. The within-tick
  recruitment cascade carries the size power law (Fig.~\ref{fig:scalefree}), whereas this per-tick spectrum
  reflects the multi-tick activity episodes (\S\ref{sec:excitable}).}
  \label{fig:powerspectrum}
\end{figure}

A globally critical, long-memory system would show a $1/f$ activity spectrum~\cite{bak1987soc}. The
canonical sandpile's flow signal is already Lorentzian ($\sim\!1/f^{2}$), not $1/f$~\cite{jensen1989onf};
genuine $1/f$ needs special conditions, such as an integrated observable with exponentially long
configuration memory~\cite{maslov1999onf}. Here the proximity graph is reshaped by motion each tick, which
keeps temporal correlations short, so we expect a Lorentzian spectrum: flat below a corner set by the
activity relaxation time, not broadband $1/f$. Measured directly, this holds
(Fig.~\ref{fig:powerspectrum}): $S_A(f)$ is a flat plateau below a corner at $f_c\approx0.03$ set by the
activity relaxation time ($\tau_{\rm ac}\approx6$ ticks, $1/(2\pi\tau_{\rm ac})\approx0.03$), then a
roll-off (slope $\approx1.3$, between $1/f$ and $1/f^2$). The plateau and corner are the signature: $A(t)$
has a \emph{finite correlation time}, a characteristic timescale, whereas a globally critical $1/f$ process
is scale-free with none. (A single slope fit to the whole curve, $\beta_{\rm spec}\approx0.6$ by detrended
fluctuation analysis, averages the flat plateau and the roll-off and is not itself the discriminator.) This
is consistent with the local-not-global reading rather than a proof of it: a short correlation time is
necessary but not sufficient for subcriticality. The matched observable bears this out---for the canonical
sandpile the \emph{integrated} mass spectrum is broadband ($\beta_{\rm mass}\approx1.9$) while the
\emph{instantaneous} activity, the observable matched to our $A(t)$, is itself nearly white
($\beta_{\rm act}\approx-0.1$).

\subsection{Active matter and foraging}
\label{sec:foraging}
The rule connects to two literatures. Its collective order is active matter of a distinct kind
(\S\ref{sec:related}); read as collective
foraging, recruitment is social area-restricted search, and blocking produces patch-leaving: crowding
drives $a/k$ below $\phi$ and releases agents, with no rate estimation, memory, or habitat average. The
behaviour is emergent patch abandonment, not a tuned marginal-value rule. Tracking patches from discovery
to depletion, they are abandoned before empty, but the giving-up density is broadly distributed (median
${\approx}40\%$, a flat distribution with few patches drained to empty) and $\phi$-independent
(Fig.~\ref{fig:patchlifecycle}), so it does not match the specific giving-up density a marginal-value rule
would predict~\cite{charnov1976optimal}. What blocking does set, tunably, is the leaving \emph{cause}: the
blocking-driven fraction of departures rises from $0$ to ${\approx}0.6$ as $\phi$ tightens $2/\phi$. At the
collective scale the same rule distributes foragers across patches, a problem with some characteristics of
the multi-armed bandit; whether it can be formalized as reinforcement learning is taken up in a companion
paper.

\subsection{How the behaviours correspond}
Each row of Table~\ref{tab:coherence} follows from the fraction rule meeting the statistics real
environments have: behaviours that foraging biology documents separately here emerge together from one
rule, wherever cheap, short-sighted searchers recruit on sparse,
patchy resources.

\begin{table}[t]\centering
\small
\renewcommand{\arraystretch}{1.3}
\begin{tabular}{@{}p{0.205\linewidth} p{0.29\linewidth} p{0.26\linewidth} p{0.085\linewidth}@{}}
\hline
\textbf{Ingredient (ours)} & \textbf{Is, classically} & \textbf{Supplied by (not tuned)} & \textbf{Shown in} \\ \hline
Cascade window (on, then off) & network contagion: LTM percolation $+$ degree-blocking~\cite{Granovetter1978,Watts2002} & the fraction rule $a/k\ge\phi$ & \S\ref{sec:results-q1} \\
Blocking (self-limitation) & patch-leaving / area-restricted search~\cite{charnov1976optimal} & the same fraction rule & \S\ref{sec:foraging} \\
Endogenous connectivity & topology--geometry feedback & mobility $+$ recruitment$\to$motion & \S\ref{sec:results} \\
Two-mode search & L\'evy foraging optimum~\cite{viswanathan2011foraging} & target sparsity $+$ clustering & \S\ref{sec:results-cost} \\
Sparse, clustered targets & generic once resources aggregate~\cite{scanlon2007} & attractive renewal, $\beta{>}0$ (max-ent field) & \S\ref{sec:generic} \\ \hline
\end{tabular}
\caption{\textbf{Each ingredient is supplied by structure already present, not fitted.} The kind of
correspondence varies: the cascade window is an exact identity; blocking-as-patch-leaving is a qualitative
correspondence (the outcome without the machinery); the L\'evy/field row is an empirical compatibility.}
\label{tab:coherence}
\end{table}

\subsection{The environment and the search: one maximum-entropy principle}
\label{sec:maxent}
Here we complete the maximum-entropy argument of \S\ref{sec:generic}, which settled the field side and left
the search's caliber basis to this section. The clumped field and the L\'evy search are
the same variational principle over different spaces. On the environment
side, maximising entropy over configurations~\cite{jaynes1957} makes the clustered field the least-committal
state once resources aggregate (\S\ref{sec:generic}). On the movement side, a truncated L\'evy walk is what
maximum entropy in its path form, Jaynes' maximum caliber~\cite{jaynes1957}, yields for movement given only
partial information~\cite{reynolds2011levy,presse2013principles}: among trajectory ensembles consistent
with limited knowledge, the heavy-tailed one is the least-committal, committing to no characteristic step
scale, mirroring a field that has none. The field's configuration ensemble $e^{\beta\mathcal{C}}$ (clumping
measure $\mathcal{C}$, attractive for $\beta{>}0$) and the path ``caliber'' $e^{-\lambda\mathcal{A}}$ make this
precise: the searcher's least-committal strategy (L\'evy) is matched to the
field's least-committal structure, visible as the optimal search exponent increasing with the field's
fractal dimension ($\alpha^\ast$ tracks $D_f$, Fig.~\ref{fig:caliberopt}; an interior regression admits
the identity $\alpha^\ast{=}D_f$ but is loosely pinned, \S\ref{sec:results-cost}). We offer this as a
conceptual correspondence, not a single partition function from which both derive; the load-bearing,
falsifiable content stays the independently-measured exponents (the forager's $\mu$ and the fields'
$\beta_{\rm res}$, \S\ref{sec:results-cost}) that were free to disagree and do not.

\paragraph{Why this rule, and what ``untuned'' means.} Our coordination rule, to activate when the active-neighbour fraction $a/k$ reaches the threshold $\phi$ and
then steer toward the recruiters, is minimal at both ends: one bit (the binary recruitment signal), one
threshold ($\phi$), and a single state-dependent switch of the step scale (recruited agents exploit locally rather than
range), which \emph{lowers}---does not add---movement per target (\S\ref{sec:results-cost}). The behaviours it produces do not
hinge on this particular rule: they follow from reward-proportional recruitment and density-dependent
self-limitation (analysed in a companion paper), not the fraction rule specifically, so soft thresholds,
weighted neighbourhoods, and other heavy-tailed laws share them, and the particular choices are
representative. ``Untuned'' does not mean parameter-free: $\phi$ and $\alpha$ set where in the
regime the swarm operates (how much blocking, how heavy-tailed), not which behaviours appear; each
holds across a range, and no behaviour was hand-fitted.

\paragraph{Status, and what remains open.} The recruitment-free single-searcher sweep (Fig.~\ref{fig:caliberopt}) shows the optimal search exponent
increasing with the field's fractal dimension ($\alpha^\ast$ tracks $D_f$; Pearson $r{=}0.97$); an
interior regression dropping the anchored endpoints still admits the identity $\alpha^\ast{=}D_f$, though
the four interior fields leave it loosely pinned (CIs in the SI). This speaks to the result that the inverse-square L\'evy
walk is not an optimal single-searcher strategy for $d\geq2$~\cite{levernier2020inverse}: that analysis
concerns encounter rate on randomly (Poisson) placed targets, where the exponent is nearly immaterial,
whereas our targets are clustered and the optimum moves with the field, the Brownian end always worst
(Fig.~\ref{fig:levyfield}). The exponent matters here precisely because the field is clustered.

Still open: per-cluster lifetime scaling; a search-theoretic L\'evy-optimality result for clustered fields
(which would put \S\ref{sec:maxent} on a formal footing); and a fair price of anarchy against a
centralized planner, which must control for information (an omniscient allocator conflates knowing-where
with coordinating), so is deferred. We should also be plain about the word general: we argue generality at
the level of the mechanism's ingredients (heavy-tailed exploration, threshold recruitment, and
density-dependent blocking on a clustered field) and demonstrate it in a single model and task family, in
simulation; we do not claim it has been shown empirically across foraging taxa, only that each ingredient
and the field statistics it needs are independently documented (\S\ref{sec:foraging}). Finally, two
deliberate scope choices. Our resources are \emph{ephemeral}: a depleted patch renews elsewhere in the
field (\S\ref{sec:methods}) at a finite renewal time, not by regrowing in place; and agents carry no
energy state or starvation, so nothing penalises failing to collect. Each isolates the recruitment
mechanism: in-place regrowth would add a temporal renewal channel, and a metabolic budget would couple
local density to survival---each a different model.

\paragraph{Beyond this model.} The rule is substrate-agnostic---it needs only cheap, short-sighted agents
recruiting on sparse, clustered opportunity---so the same coupling is a candidate wherever those
ingredients recur, including in three and higher dimensions. Abstractly it sits near swarm metaheuristics
(particle-swarm and kin), with clustered optima playing the role of the resource field, and near
reinforcement-learning-style allocation without a learner (a companion
paper~\cite{wilkerson2026learnerfree}); as a social contagion (its linear-threshold lineage) it suggests
reward-driven herding in markets and adoption dynamics. Concretely it maps onto engineered
collectives---sensing or search-and-rescue drone swarms, and software agent teams coordinating over sparse
information---all directions to test, not results.

\section{Conclusion}
\label{sec:conclusion}
Cascade-mediated recruitment is a minimal coordination primitive: percolation enables it and blocking
self-limits it (the linear-threshold cascade window in geometric space), producing an emergent
exploration--exploitation balance. The self-limitation is caused by the topology--geometry coupling; the swarm self-organizes to
quasi-criticality (a near-critical operating point, below the transition, that never settles onto a clean
critical state, of the non-conservative SOC family), with an activity spectrum that has a finite correlation
time rather than the $1/f$ of global criticality. Coordination is efficient in the sparse,
clumped fields that dominate real foraging, where one threshold rule reproduces, together,
recruitment, patch-leaving / area-restricted search, and L\'evy-then-local
search. The resource field and the search are both least-committal---the field a clustered maximum-entropy process,
the search a maximum-caliber walk---and the optimal search exponent
tracks the field's fractal dimension, so the coordination is matched to generic environment statistics
rather than fine-tuned.

\section*{Competing interests}
The author declares a competing financial interest: a U.S.\ provisional patent application has been filed
(assigned to Komplexity AI LLC) covering aspects of the coordination method described.

\section*{Data and code availability}
Complete simulation code and figure-regenerating scripts are at
\url{https://github.com/galenwilkerson/collective-foraging-cascades} (publicly available); the
sweep data regenerates from the code (or point the \texttt{FLOCKING\_DATA} environment variable at a
directory of cached \texttt{.npz} files).

\bibliographystyle{aipnum4-2}
\bibliography{refs}

\clearpage
\onecolumngrid
\setcounter{figure}{0}
\setcounter{table}{0}
\setcounter{equation}{0}
\renewcommand{\thefigure}{S\arabic{figure}}
\renewcommand{\thetable}{S\arabic{table}}
\renewcommand{\theequation}{S\arabic{equation}}
\renewcommand{\topfraction}{0.95}
\renewcommand{\bottomfraction}{0.95}
\renewcommand{\textfraction}{0.05}
\renewcommand{\floatpagefraction}{0.70}
\renewcommand{\dbltopfraction}{0.95}
\renewcommand{\dblfloatpagefraction}{0.70}
\setcounter{topnumber}{4}
\setcounter{bottomnumber}{4}
\setcounter{totalnumber}{8}
\setcounter{dbltopnumber}{4}
\setlength{\textfloatsep}{10pt plus 2pt minus 2pt}
\setlength{\floatsep}{10pt plus 2pt minus 2pt}
\setlength{\dblfloatsep}{10pt plus 2pt minus 2pt}
\vspace{1em}
\begin{center}{\large\textbf{Supplementary Information}}\end{center}
\vspace{0.5em}
\twocolumngrid

\section*{Supplementary figures and analyses}
\label{app:si}
These support the genericity (\S\ref{sec:generic}), Q2 criticality (\S\ref{sec:results-q2}), cost and
field-adaptation ($\alpha^\ast\!\approx\!D_f$, \S\ref{sec:results-cost}), balance (\S\ref{sec:excitable}), and foraging
(\S\ref{sec:foraging}) results; placed here to keep the main text focused.

\paragraph{How far the maximum-entropy reading reaches (\S\ref{sec:maxent}).} The maximum-entropy reading
covers the field and the search, where imposing a constraint derives the law. It does not cover the
cascade: the cascade's near-criticality is self-organized, appearing only near the percolation point, and
whether such behaviour admits a maximum-caliber account~\cite{dewar2003,presse2013principles} is a
long-standing open question~\cite{jensen1998soc,moraBialek2011,schwab2014} we do not settle here.

\begin{figure*}[tb]\centering
  \includegraphics[width=\linewidth]{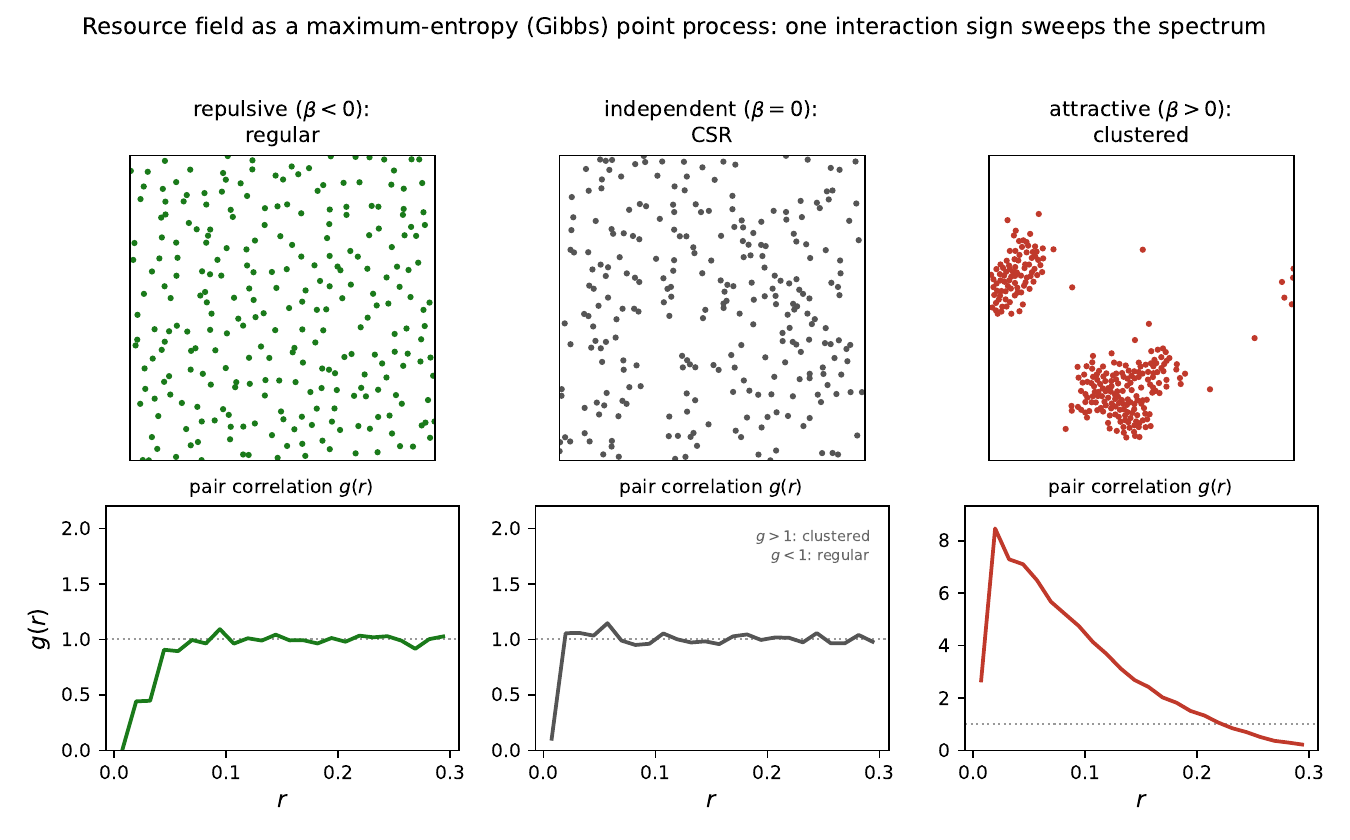}
  \caption{\textbf{The resource field as a maximum-entropy (Gibbs) point process: one interaction sign sweeps
  the spectrum.} Samples of $P\propto e^{\beta\mathcal{C}}$ ($\mathcal{C}$ the number of close pairs).
  \textbf{Top:} point patterns for repulsion ($\beta{<}0$, regular), independence ($\beta{=}0$, CSR), and
  attraction ($\beta{>}0$, clustered). \textbf{Bottom:} the corresponding pair correlation $g(r)$ (the neighbour density at separation $r$ relative to a completely random field: $g{>}1$ clustering, $g{<}1$ regularity, $g{=}1$ randomness), below $1$
  at short range for the regular field, flat ($\approx1$) for CSR, and peaked above $1$ for the clustered
  field. Regularity requires an active repulsion and is the exception; the generic, attractive
  resource-renewing interaction makes the clustered field the least-committal one (\S\ref{sec:generic}).}
  \label{fig:entropy}
\end{figure*}

\paragraph{The resource-field spectrum (\S\ref{sec:generic}).} Fig.~\ref{fig:entropy} samples the
maximum-entropy (Gibbs) point process $P\propto e^{\beta\mathcal{C}}$ ($\mathcal{C}$ the number of close
pairs) across the sign of the interaction: repulsion ($\beta{<}0$) gives a regular,
fluctuation-suppressed field (pair correlation $g(r){<}1$ at short range); independence ($\beta{=}0$)
gives complete spatial randomness ($g(r){\approx}1$); attraction ($\beta{>}0$) gives a clustered field
($g(r){>}1$, peaked). Regularity is the exception, requiring an active repulsion, while the generic
resource-renewing interaction is attractive, so the least-committal field is clustered. The regular limit's
cost is explicit as a partition function: for $M$ targets in $K{=}M$ cells the perfectly regular state has
weight $M!/M^M\sim e^{-M}$ and is reached only as $\beta\to-\infty$, at an extensive free-energy cost
$\ln(M^M/M!)=M$ nats, an entropic force away from regularity.

\begin{figure}[htbp]
  \centering
  \includegraphics[width=\linewidth]{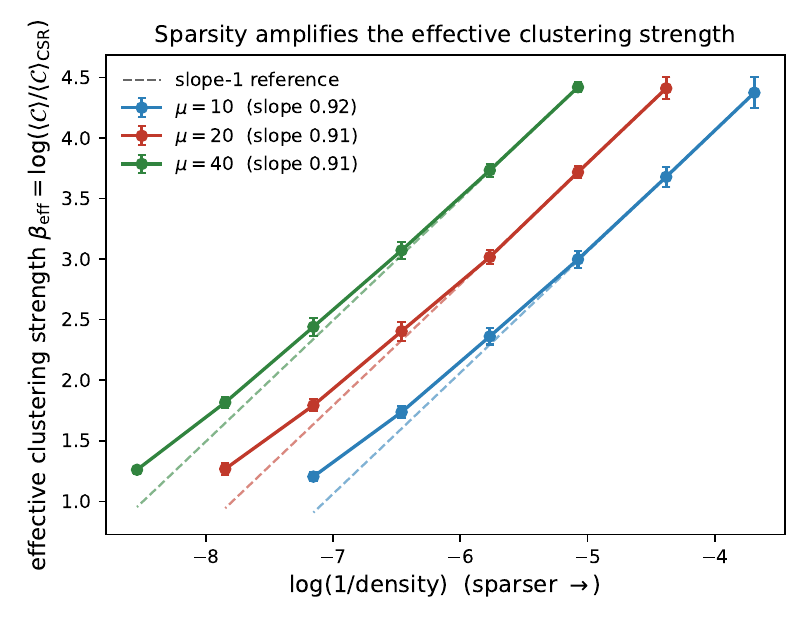}
  \caption{\textbf{Sparsity amplifies the effective clustering strength.}
  $\beta_{\rm eff}=\log(\langle\mathcal{C}\rangle/\langle\mathcal{C}\rangle_{\rm CSR})$ for a Thomas cluster
  field versus $\log(1/\text{density})$, at fixed clump occupancy $\mu$ (points: mean${\pm}$s.d.\ over $24$
  realizations; dashed: a slope-$1$ reference anchored at each series' densest (leftmost) point). The slope is
  ${\approx}0.91$ and $\mu$-independent, as the derivation requires, while the intercept rises with $\mu$.
  Sparser fields (right) carry larger effective clustering: the sparsity foraging presupposes sharpens the
  structure recruitment exploits.}
  \label{fig:betasparsity}
\end{figure}

\paragraph{Sparsity amplifies the effective clustering (\S\ref{sec:generic}).} The effective clustering
strength of a field is the dilute-limit multiplier that reproduces its mean close-pair count,
$\beta_{\rm eff}=\log(\langle\mathcal{C}\rangle/\langle\mathcal{C}\rangle_{\rm CSR})$, where
$\langle\mathcal{C}\rangle_{\rm CSR}=\binom{N}{2}\pi r_0^2$ is the count a CSR field of the same density $N$
gives within range $r_0$. For a renewal (cluster) field of $N_p$ clumps with occupancy $\mu$ (total
$N{=}N_p\mu$), within-clump pairs dominate: $\langle\mathcal{C}\rangle\approx N_p\binom{\mu}{2}\approx
\tfrac12 N\mu$, linear in $N$, whereas $\langle\mathcal{C}\rangle_{\rm CSR}\approx\tfrac12 N^2\pi r_0^2$ is
quadratic. Hence $\beta_{\rm eff}\approx\log\mu-\log N-\log(\pi r_0^2)$: at fixed occupancy $\mu$ it rises
like $\log(1/\text{density})$ with unit slope. Fig.~\ref{fig:betasparsity} confirms this on a Thomas cluster
process (Gaussian offspring at scale $\sigma_c$, $r_0{\ge}\sigma_c$): the fitted slope is $0.91$--$0.92$,
independent of $\mu\in\{10,20,40\}$, with the intercept shifting by $\log\mu$ as predicted (the small
constant offset below the idealized value is the finite offspring spread, which leaves some within-clump pairs
just beyond $r_0$). Two caveats bound the claim: the growth is logarithmic, and it amplifies
contrast ($g(r)$, hence $\beta_{\rm eff}$), not the absolute windfall per discovery, which also
depends on the occupancy $\mu$ that renewal dynamics set. It is also distinct from the fluctuation clumpiness
of a sparse CSR field, where $g(r){=}1$: here the excess pair correlation is genuine ($g(r){\gg}1$), and it
is its measured magnitude that sparsity amplifies.

\paragraph{Cost anatomy and communication channels (\S\ref{sec:results-cost}).} The two communication
channels must be separated. Presence beacons are one per agent per tick regardless of behaviour
(constant, one per agent, ${\approx}N$/tick; rank correlation $-0.001$ with $z$), a clock, not a signal, paid
identically by both conditions, so total bits is $93\%$ rank-determined by time-to-completion. The
recruitment-specific activation channel is density-dependent ($+0.42$ with $z$), tracing the same
inverted-U (rising $21\to91$ pings/tick, falling to $86$ as blocking sets in). On that channel recruitment
costs ${\sim}4$--$5\times$ more messages ($0.254,0.249,0.245$ at $\alpha=1,1.5,2$); since activation is one
ping per active agent, the ratio is cascade amplification restated, governed by $z$ and $\phi$
(${\sim}2\times$ at $z{\approx}1$ to ${\sim}12\times$ at $z{\approx}3$--$13$), flat in $\alpha$.

\begin{figure*}[tb]\centering
  \includegraphics[width=\linewidth]{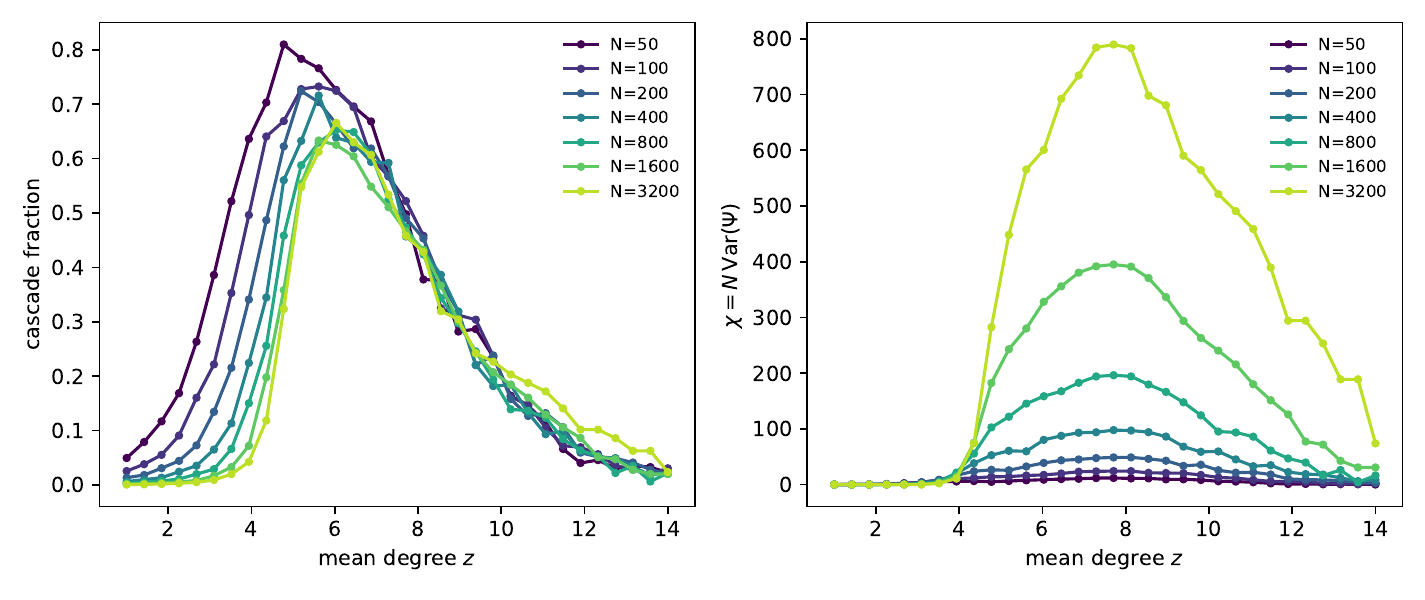}
  \caption{\textbf{Finite-size scaling pins a size-independent critical point.} \textbf{(a)} Cascade
  fraction (largest active-cluster fraction $\Psi$) vs.\ $z$ for each $N$. The curves are not
  $N$-independent; they undergo finite-size scaling: small systems round the onset below the RGG
  percolation threshold $z_c^{\rm RGG}\approx4.51$ and depress the peak, sharpening toward a fixed window as
  $N$ grows, and cross near $z^\ast\approx7.8$. What is $N$-independent is the critical point, not the
  finite-$N$ curves. \textbf{(b)} That point is pinned by the susceptibility
  $\chi=N\,\mathrm{Var}(\Psi)$ ($\Psi$ largest-cluster fraction), whose peak stays at fixed
  $z^\ast\approx7.8$ for all $N\in[50,3200]$ (location s.d.\ $0.15$) with height $\propto N$.}
  \label{fig:fss}
\end{figure*}

\paragraph{Universality (finite-size scaling).} Sweeping $z$ for $N\in\{50,\dots,3200\}$ (a $64\times$ range), the susceptibility
$\chi=N\,\mathrm{Var}(\Psi)$ ($\Psi$ the largest-cluster fraction) peaks at a fixed dimensionless $z^\ast\approx7.8$ for every $N$ (location s.d.\
$0.15$), peak height $\propto N^{1.0}$, the window sharpening to this $N$-stable critical point as
$N$ grows (Fig.~\ref{fig:fss}): the
cascade transition is a property of the dimensionless control, not system size. (The single-seed FSS setup
gives a lower absolute cascade fraction than Fig.~\ref{fig:invertedU}'s fixed-fraction seeding; what is
$N$-independent is the window, not the peak height. $z^\ast\approx7.8$ sits just above the
cascade-fraction peak $z\approx6$--$7$ because a susceptibility maximum generically lies above where the
order parameter turns over.)

\begin{figure*}[tb]\centering
  \includegraphics[width=\linewidth]{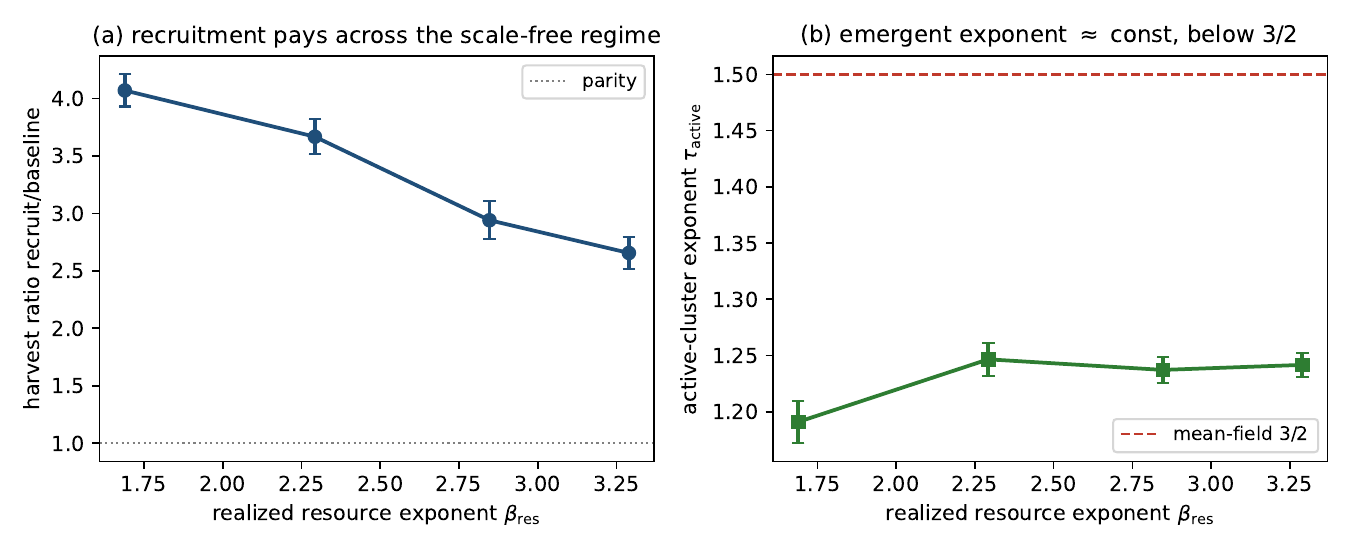}
  \caption{\textbf{On power-law resource fields, the payoff tracks the field, and the cascades stay near-critical
  at a field-set exponent.} \textbf{(a)} Harvest ratio $>1$ across the whole scale-free regime, rising as
  the field clumps (smaller $\beta_{\rm res}$); the benefit is field-dependent. \textbf{(b)} The emergent
  cascade exponent $\tau_{\rm active}$ is roughly constant within the scale-free family
  ($\approx1.2$, range $1.19$--$1.25$, across the resource-exponent range $\beta_{\rm res}$, no significant trend,
  overlapping $95\%$ CIs), but sits below the near-CSR value (median $\tau\approx1.5$, itself a
  near-critical band, \S\ref{sec:results-q2}): the cascades stay near-critical across these fields while the
  exponent is set by the clustered regime, not a constant the swarm imposes. Both panels: markers,
  connecting line, and shaded $95\%$ bootstrap-CI band.}
  \label{fig:scalefree}
\end{figure*}

\begin{figure}[htbp]\centering
  \includegraphics[width=\linewidth]{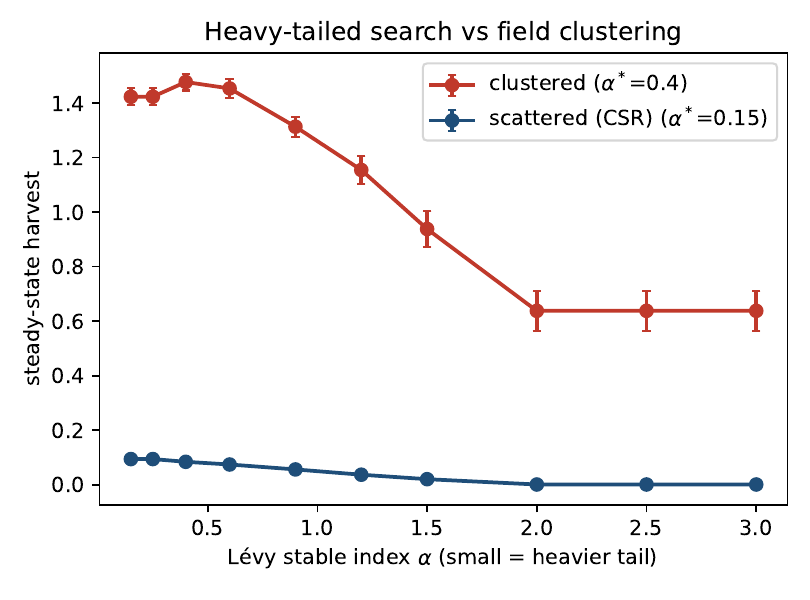}
  \caption{\textbf{Heavy-tailed search is advantageous for clustered targets too.} Steady-state harvest
  vs.\ $\alpha$ (smaller $=$ heavier tail) for a clustered ($3$ clumps) vs.\ a scattered (CSR/Poisson) field
  harvest peaks
  at a heavy tail for both ($\alpha^\ast\!\approx\!0.4$ clustered, $\alpha^\ast\!\approx\!0.15$ scattered), the clustered field paying several-fold more. Bands: $95\%$ bootstrap CI
  over replicas. (Fig.~\ref{fig:caliberopt} quantifies this over a continuous range of scale-free fields.)}
  \label{fig:levyfield}
\end{figure}

\paragraph{Scale-free resource fields and L\'evy-on-clustered.} Driving the model with power-law clump
sizes (realized $\beta_{\rm res}\in\{1.7,2.3,2.8,3.3\}$), recruitment still pays off
($2.7$--$4.1\times$ baseline, rising as the field clumps), while the emergent active-cluster exponent is roughly constant within the scale-free family:
$\tau_{\rm active}\approx1.2$ (range $1.19$--$1.25$) across the whole $\beta_{\rm res}$ range (overlapping $95\%$ CIs, no
significant trend), though this sits below the near-CSR value (median $\tau\approx1.5$, itself a near-critical band)
(\S\ref{sec:results-q2}). Blocking keeps the cascades near-critical on clustered fields, but at a lower,
geometry-set exponent than the near-CSR case: the criticality is robust across fields while the exponent is
set by the field, not a constant the swarm imposes (Fig.~\ref{fig:scalefree}). Sweeping $\alpha$ on clustered vs.\
scattered fields, harvest peaks at a heavy tail ($\alpha\lesssim1$) for both, the clustered field
paying several-fold more (Fig.~\ref{fig:levyfield}), so heavy-tailed search is advantageous for
clustered targets too, as the maximum-entropy reading predicts. An independent, recruitment-free
single-searcher check (promoted to the main text, Fig.~\ref{fig:caliberopt}) quantifies this: on scale-free
L\'evy-dust fields the optimum $\alpha^\ast$ tracks the field's fractal dimension, $\alpha^\ast\!\approx\!D_f$
over the clustered range, the Brownian end always worst (Pearson $r{=}0.97$ on the fitted optima,
$r{\approx}0.86$ by raw argmax; an interior regression dropping the anchored ballistic and inverse-square
endpoints gives bootstrap slope $1.18$, 95\% CI $[0.91,1.54]\ni1$, intercept CI $[-0.42,0.17]\ni0$---the
identity $\alpha^\ast{=}D_f$ admitted but loosely pinned by the four interior fields).

\begin{figure*}[tb]\centering
  \includegraphics[width=\linewidth]{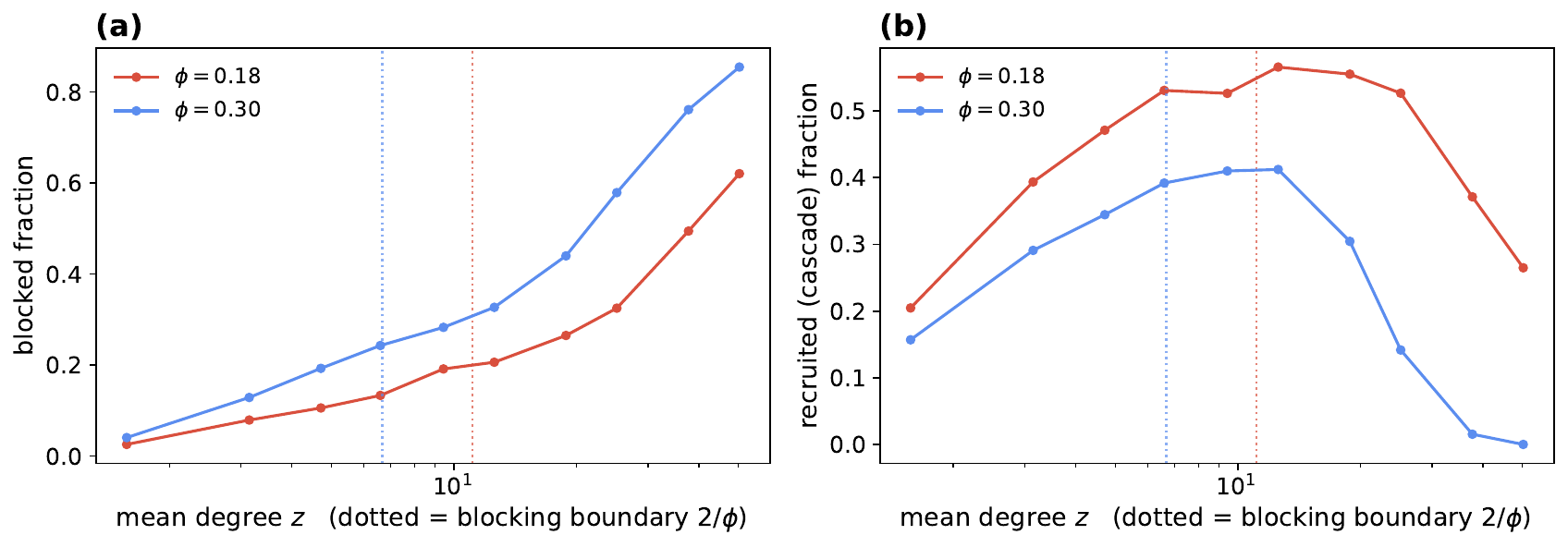}
  \caption{\textbf{Blocking is the brake at high density.} \textbf{(a)} The blocked fraction rises with
  mean degree $z$, faster for larger $\phi$. \textbf{(b)} The recruited fraction peaks in the window
  then collapses; dotted lines mark $z=2/\phi$.}
  \label{fig:blockdep}
\end{figure*}

\paragraph{Blocking measured directly.} Fig.~\ref{fig:blockdep}: the blocked fraction (agents with active
neighbours but $a/k<\phi$) grows monotonically with mean degree ($0.03\to0.62$ at $\phi{=}0.18$,
$0.04\to0.85$ at $\phi{=}0.30$) and is stronger at larger $\phi$ (lower boundary $2/\phi$), while
the recruited fraction traces the inverted-U, confirming blocking as the operative brake.

\begin{figure*}[tb]
  \centering
  \includegraphics[width=\linewidth]{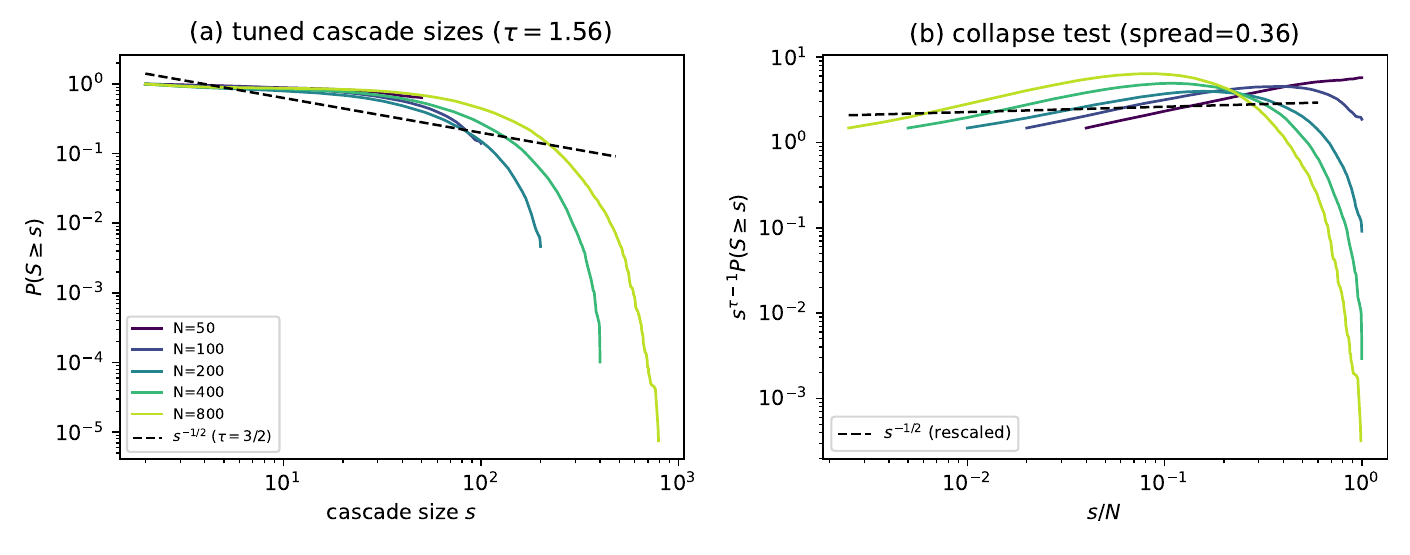}
  \caption{\textbf{Finite-size scaling: even tuned, the cascades sit in a near-critical band, not a clean
  mean-field collapse.} At matched criticality ($\sigma_{\rm aval}{\approx}1$, the activation threshold calibrated
  per $N$; legend) on the near-CSR (maximum-entropy) field. \textbf{(a)} Survival distributions
  $P(S{\ge}s)$ for $N{=}50$--$800$: the cutoff grows with the swarm size, shown against the mean-field reference
  $s^{-1/2}$ (dashed). \textbf{(b)} The finite-size-scaling test: $s^{\tau-1}
  P(S{\ge}s)$ against $s/N$ does \emph{not} overlay onto a single curve---the rescaled survival functions fall in
  a band rather than collapsing onto the mean-field reference ($\tau{=}3/2$, $s^{-1/2}$). So even a deliberately tuned transition is a near-critical band rather
  than a clean mean-field collapse, consistent with the self-organized quasi-criticality of the untuned swarm (\S\ref{sec:results-q2}).}
  \label{fig:fsscollapse}
\end{figure*}

\begin{figure*}[tb]\centering
\begin{minipage}[t]{0.49\textwidth}\centering
  \includegraphics[width=\linewidth]{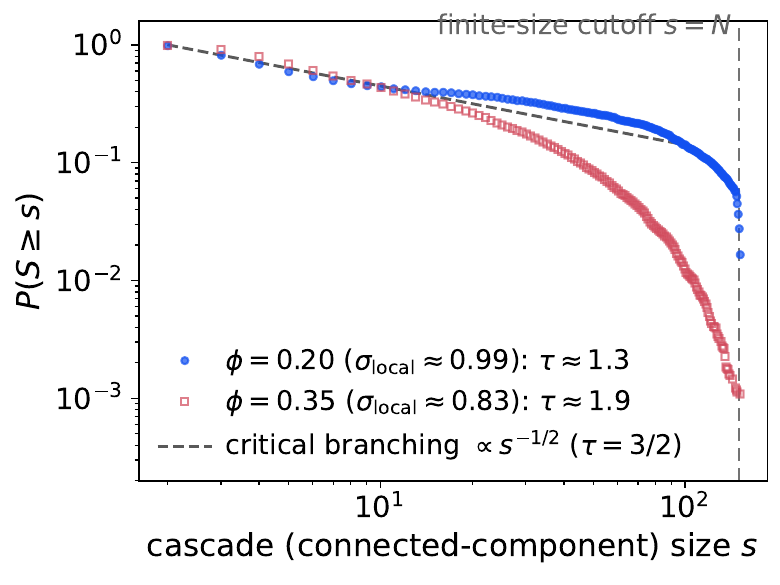}
  \caption{\textbf{Cascade-size distributions at two thresholds.} Survival distribution $P(S{\ge}s)$ of
  cascade size $s$ (connected-component size of one branching cascade in the active subgraph of the
  radius-$r$ proximity graph; L\'evy $\alpha{=}1.5$, on the swarm's clustered operating field), at two
  activation thresholds. At $\phi{=}0.20$ ($\sigma_{\rm local}{\approx}0.99$, blue) cascades span a
  truncated power law up to the finite-size cutoff $s{=}N{=}150$ (dashed vertical), slope
  $\tau{\approx}1.3$, shallower than the mean-field reference $P(S{\ge}s)\propto s^{-1/2}$ ($\tau{=}3/2$,
  dashed); the mean-field value is approached only under the tuned collapse on the near-CSR field
  (Fig.~\ref{fig:fsscollapse}). At $\phi{=}0.35$ ($\sigma_{\rm local}{\approx}0.83$, red) the tail bends
  into an exponential ($\tau{\approx}1.9$). The fitted exponent is bounded by the $s{\le}N$ cutoff and
  sensitive to the fit window, so $\sigma_{\rm local}$, not this slope, is the branching measure.}
  \label{fig:recruitcluster}
\end{minipage}\hfill
\begin{minipage}[t]{0.49\textwidth}\centering
  \includegraphics[width=\linewidth]{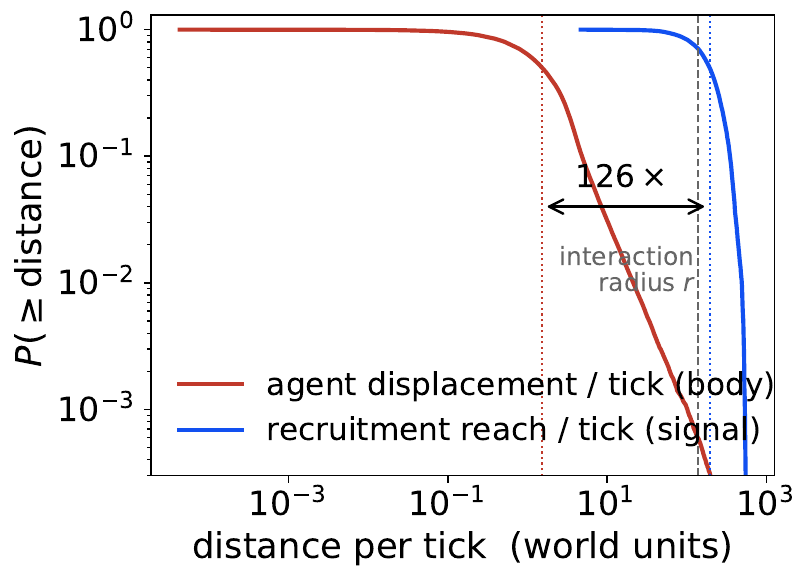}
  \caption{\textbf{Recruitment reaches across the swarm in one cascade.} Survival distributions, per tick,
  of the recruitment reach (blue; distance from a newly activated agent to the nearest previously active
  one) and the agent displacement (red). The reach distribution sits well above the displacement
  distribution (dotted medians; double arrow); dashed line marks the interaction radius $r$. This is a
  per-cascade recruitment reach against a locomotion step, not a physical front speed: coordination
  travels along the percolating network, which spans distances an agent crosses only over many ticks.}
  \label{fig:frontspeed}
\end{minipage}
\end{figure*}

\paragraph{The tuned transition is a near-critical band (\S\ref{sec:results-q2}).} Tuning the threshold to
$\sigma_{\rm aval}{\approx}1$ per $N\in\{50,\dots,800\}$ on the near-CSR field, the cascade-size
distributions do \emph{not} collapse onto a single clean curve but fall in a near-critical band
(Fig.~\ref{fig:fsscollapse}): rescaled by the median $\tau\approx1.56$, the curves fall in a band rather than
overlaying.
Even a deliberately tuned transition thus does not settle onto a clean, size-invariant critical state; the
size-independent critical point is pinned instead by the order-parameter susceptibility above.

\begin{figure*}[tb]\centering
  \includegraphics[width=\linewidth]{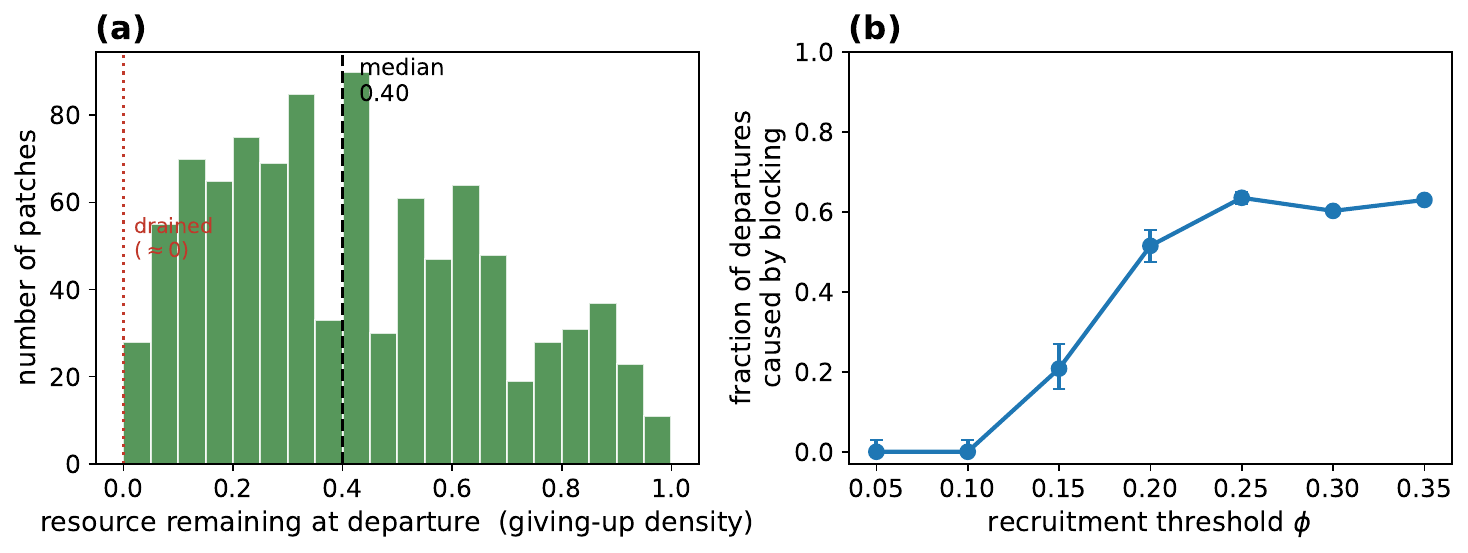}
  \caption{\textbf{Patches are abandoned before depletion, and blocking is the measured cause.}
  \textbf{(a) Left before empty.} Giving-up density (resource remaining when a patch's occupancy
  collapses from its peak) broadly distributed with median near $0.40$ (a flat distribution, not a sharp
  peak), few patches drained to empty: patches are abandoned before drained (area-restricted-search-like),
  without the tuned giving-up density a marginal-value rule predicts. \textbf{(b) Blocking drives leaving.} The
  blocking-driven fraction of departures (an agent squeezed below threshold while it still has active
  neighbours) rises with the recruitment threshold $\phi$: crowding, tunably, releases agents. Each point
  is the fraction of $n{=}617$ departures caused by blocking; bars are the Wilson $95\%$ confidence
  interval on that proportion.}
  \label{fig:patchlifecycle}
\end{figure*}

\paragraph{Patch lifecycle: leaving before depletion (\S\ref{sec:foraging}).} Tracking patches from
discovery to abandonment, the giving-up density is broadly distributed and $\phi$-independent, while the
blocking-driven fraction of departures rises with $\phi$ (Fig.~\ref{fig:patchlifecycle}): blocking is the
measured leaving cause, an emergent patch abandonment rather than a tuned marginal-value rule.

\begin{figure*}[tb]\centering
\begin{minipage}[t]{0.49\textwidth}\centering
  \includegraphics[width=\linewidth]{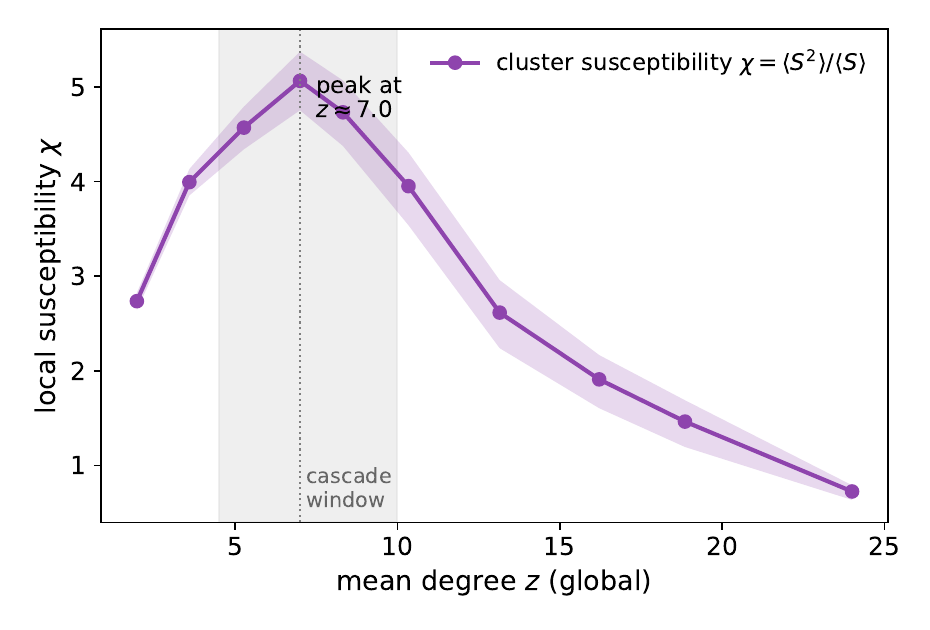}
  \caption{\textbf{A corroborating percolation signature: local susceptibility peaks inside the cascade
  window.} $\chi=\langle S^2\rangle/\langle S\rangle$ (largest cluster excluded) vs.\ mean degree $z$, for
  the swarm's own L\'evy exploration ($\alpha{=}1.5$), with cascade fraction (dashed): a broad peak at
  $z\approx7.0$ (within the cascade window, below the finite-size-scaling $z^\ast\approx7.8$), the
  percolation-susceptibility hallmark. Shaded band: $95\%$ bootstrap CI over replicas.}
  \label{fig:susceptibility}
\end{minipage}\hfill
\begin{minipage}[t]{0.49\textwidth}\centering
  \includegraphics[width=\linewidth]{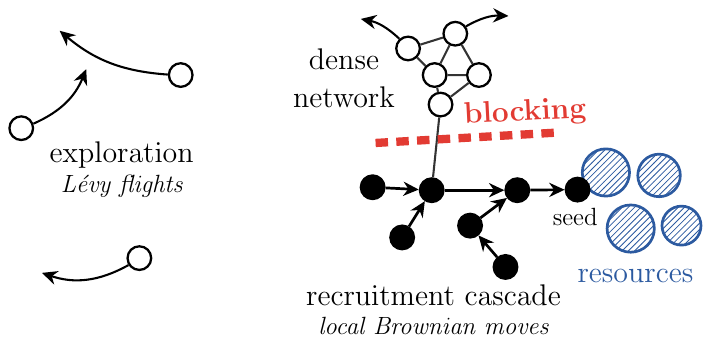}
  \caption{\textbf{Spatial schematic of the mechanism's phases: explore, recruit to a discovery, then self-limit.}
  Agents explore by long-range (L\'evy) search; a discovery seeds a linear-threshold recruitment
  cascade on the proximity graph, drawing neighbours in, while density-dependent blocking
  self-limits it and releases agents back to exploration. (Referenced from the main text as the spatial
  picture of the cycle of Fig.~\ref{fig:cycle}.)}
  \label{fig:overview}
\end{minipage}
\end{figure*}

\clearpage
\end{document}